\documentclass[12pt]{JHEP3}

\usepackage{amsmath}

\title{Entropy Current from Partition Function: One Example}
\author{  Sayantani Bhattacharyya\\ 
Indian Institute of Technology Kanpur, Kanpur, India-208016\\
email: sayanta@iitk.ac.in
}

\abstract{In hydrodynamics the existence of an entropy current with non-negative divergence is 
related to the existence of a time-independent solution in a static background. Recently there has been a proposal for how to construct an entropy current from the equilibrium  partition function of the fluid system. In this note, we have applied this algorithm for the charged fluid at second order in derivative expansion. From the partition function we first constructed one example of entropy current with non-negative divergence upto the required order. Finally we extended it to its most general form, consistent with the  principle of local entropy production. As a by-product we got the constraints on the second order transport coefficients for a parity even charged fluid, but in some non-standard fluid frame.}
\preprint{}

\begin{document}

\section{Introduction}\label{sec:intro}
Fluid dynamics is an effective description for near equilibrium physics. In the `fluid limit', it is possible to describe the system only by a few classical functions (much fewer than the number of degrees of freedom, the underlying microscopic theory has). Because it is a system, only slightly away from equilibrium, it is assumed that each of these functions (or fluid variables)  vary slowly in time and space, so that the number of space-time derivatives can be treated as an expansion parameter.
\newline
Equations of fluid dynamics are just the conservation equations for the stress tensor and other conserved currents of the system.
The basic input here are the constitutive relations where the conserved quantities are expressed in terms of a few fluid variables like velocity, temperature, chemical potentials etc. These constitutive relations  are always expanded in terms of the derivatives of the fluid variables. Each independent term in the constitutive relations is multiplied by some coefficient that controls the transport properties in the fluid. Once the values for these coefficients (called transport coefficients) are specified, the system of equations for a given fluid is completely fixed.

 Generically it is very difficult to determine the details of a system in the `fluid limit', starting from its microscopic description. In other words, it is practically impossible to  compute the transport coefficients for a strongly correlated system in its fluid limit. Therefore, as it is usual for any effective theory,  the important question here is to figure out the most general structure of the fluid equations, considering only
some universal aspects or laws of nature that any physical system  must obey. We do not yet know the exhaustive list of such principles that we should impose on the equations of fluid dynamics. However we know some of them and we often observe an overlap between the constraints generated from two different sets of physical requirement.
\newline
Ultimately we would like to determine a final theory of fluid dynamics where all such consistency conditions are automatically taken care of and where we would have a genuine count for the number of independent transport coefficients to be fixed only by some experiments. If we have this goal in mind, it is very important to explore the consequences of imposing the laws we already know and in particular the details of their overlap or equivalence in this context. 

The `existence of stable equilibrium' and `local entropy production in every non-equilibrium flow' are two such important universal physical requirements, which any set of fluid equations must satisfy. It turns out that these two conditions not only  constrain the fluid equations to a large extent, but also independently end up almost in the same set of constraints at every order. 
There are several  examples where this equivalence have been worked out for different systems(see for instance \cite{oddloga},\cite{oddamos},\cite{superfluident},\cite{equipart},\cite{equipartamos},\cite{equiloga},\cite{equisuperfluid},\cite{rindlerhydro}). But in all these examples, the constraints were computed separately by imposing both the conditions (i.e. existence of equilibrium and local entropy production) and the overlap was observed in some `experimental sense', without knowing why this happens or how universal it is.
\newline
Let us explain this point in more detail.  In these particular systems (as described in \cite{oddloga},\cite{oddamos},\cite{superfluident},\cite{rindlerhydro},\cite{Romat},\cite{secondorder}) the local entropy production was proved by  explicit construction of the most general entropy current with non-negative divergence.  It was shown that the construction was possible only if the fluid equations satisfy some constraints, which (except few inequalities) are same as the ones derived from the existence of equilibrium.  But there were no universal algorithm for this construction of entropy current. We often need to use clever argument, very specific to the system being studied. In particular, it was not clear whether some properties of equilibrium could be used to contruct the entropy current, thus revealing the reason behind the overlapping constraints, arising from these two conditions.
\newline
In \cite{entpart} we have explored the reasons behind such overlap to all orders in derivative expansion. We have been able to formulate an algorithm to construct one example of entropy current with non-negative divergence using the partition function of the system. Moreover, in \cite{entpart} it has been claimed that if we assume the existence of a stable equilibrium\footnote{Existence of a stable equilibrium imposes few inequalities only on the first order transport coefficients and reduce the number of independent coefficients at every order by relating them to each other.} in any static background, then the algorithm described for the construction of the entropy current will always work, at any given order in derivative expansion.
\newline
In this note, our goal is to test this algorithm (as proposed in \cite{entpart}) for a sufficiently complicated system of charged fluid at second order in derivative expansion, though we shall restrict ourselves only to the parity even sector\footnote{In particular, our analysis is insensitive to any anomaly that a general fluid might have. For our case the stress tensor and the current will be exactly conserved.}. Here we shall first determine the entropy current of the system  from its partition function using the algorithm as described in \cite{entpart}. Next we shall explicitly compute the divergence of the entropy current and shall show that it is indeed non-negative, thus verifying the claims of \cite{entpart} for this complicated but explicit case.

Though the main purpose of this note is just to show how the algorithm described in \cite{entpart} works out, this analysis will also implicitly generate a new set of physical constraints to be imposed on the transport coefficients of charged fluid at second order in derivative expansion. This set of constraints itself could be important in the context of high energy physics. For example, in RHIC and LHC heavy ion experiments, the expansion of the hot and dense plasma is often modelled by fluid equations including second order corrections. Since we do not know how to compute the properties of this fluid from the underlying theory of QCD, it would be useful to have a classification of all the allowed transport coefficients of any charged fluid at second order.
\newline
Another context where these set of constraints might turn out to be useful is in the literature of `fluid-gravity correspondence'. Transport coefficients for some conformal charged fluids have been exactly computed using their gravity duals (for example the holographic computation as done in \cite{chargedbrane}, \cite{chargeAmos}, \cite{holography}). The constraints we derived here must be obeyed by these fluids.\footnote{ These papers used a different fluid-frame than the one we have used. The frame that we have used is very well-suited for this analysis using the equilibrium parttion function, though this is not a standard frame to be used in any relativistic fluid literature. Had our purpose been to constrain the  transport coefficients of the second order charged fluid, we would have translated our results to the standard Landau or Eckert frame by redifining our fluid variables appropriately. Then only we would have been able to compare our results with what found in literature.} Similar analysis, using the existence of the partition function, has already been done for 2nd order charged fluid in the parity odd sector (see for instance \cite{conformalhydro},\cite{equijustin}). But here we shall restrict ourselves just to the implementation of the algorithm as described in \cite{entpart}. We shall leave the full analysis of the charged fluid for future work.

Now in the rest of this section we shall briefly describe the results we found in this note.
 It turns out that for parity even charged fluid at second order in derivative expansion the most general equilibrium partition function (on any static background with curvature small enough to be expanded in terms of derivatives) could have 7 parameters that are free functions of local temperature and the chemical potential. Therefore the non-dissipative part of the stress tensor and the current( the part that does not vanish in equilibrium) is completely determined in terms of these seven parameters. The entropy current that we have determined from the partition function also has these same seven parameters. Apart from these coefficients the entropy current could also have 10 free functions which are not determined by the partition function. We have called these free parameters as `ambiguity' in the prescription we used to determine the entropy current. But this is not a contradiction since using the algorithm (as described in \cite{entpart}) we can construct one example of the entropy current with non-negative divergence upto the given order. There is no claim for uniqueness.
 
 So finally we have determined the most general possible entropy current that would be consistent with the requirement of local entropy production. It has 17 parameters. As mentioned before in all our calculation we worked in a very non-standard half-fixed fluid frame. Our choice of frame is well-defined when the system is in equilibrium. But outside equilibrium we have assumed the most general possible extension of this frame without fixing to anything particular\footnote{Since we have restricted ourselves to parity even fluid without any anomaly, it turns out that the expression for the second order corrections to the entropy current (the `non-canonical' part) is frame-invariant. This is a very special feature of non-anomalous fluids at second order. The main reason is that in this case the second order correction is actually the leading correction in terms of derivative expansion}
 \newline
  Below we are quoting the final expression of the most general entropy current at second order in derivative expansion. 
 \begin{equation}\label{finalresult1}
 \begin{split}
 \text{Entropy current}=J^\mu =&~ J^\mu_{can} + S^\mu + S^\mu_{zero-divergence} + S^\mu_B\\
 \end{split}
 \end{equation}
 Here $J^\mu_{can}$ is the canonical part of the entropy current defined in terms of the `non-ideal' part of the stress tensor and the current. $S^\mu$ is the piece that is determined using the partition function. $S^\mu_{zero-divergence}$ ans $S^\mu_B$ together captures the `ambiguity' i.e. the terms that could not be fixed from the principle of local entropy production alone. Below we are giving explicit expressions for each of these four terms.
 \begin{equation}\label{finalresult2}
 J^\mu_{can} = s u^\mu - \frac{u_\nu\pi^{\mu\nu}}{T}-\nu j^\mu
 \end{equation}
 Here $s$ is the entropy density of the system. $u^\mu$, $T$ and $\nu$ denote the velocity, temperature and  the chemical potential  of the fluid respectively. $\pi^{\mu\nu}$ and $j^\nu$ collectively denote all the derivative corrections to the stress tensor and the charge current respectively.
 \begin{equation}\label{finalresult3}
 \begin{split}
 S^\mu =&\sum_{i=1}^7 S^\mu_{(i)}\\
S^\mu_{(1)}&= K_T \left[(D_\nu T)(D^\nu T)u^\mu -2(D^\mu T)(u.\partial T)\right]\\
S^\mu_{(2)}&=K_c \left[(D_\nu \nu)(D^\nu \nu)u^\mu -2(D^\mu \nu)(u.\partial T)\right]\\
S^\mu_{(3)}&=K_{cT} \left[(D_\nu \nu)(D^\nu T)u^\mu -(D^\mu \nu)(u.\partial \nu)-(D^\mu \nu)(u.\partial T)\right]\\
S^\mu_{(4)}&=4T^2K_f\left[( \omega_{ab}\omega^{ab})u^\mu- 2\omega^{\mu\nu} h_\nu\right]\\
S^\mu_{(5)}&=K_{Ff}\left[-2T(H_{ab}~\omega^{ab})u^\mu +2T H^{\mu\nu} h_\nu-4T^2 \omega^{\mu\nu}(V_\nu -\nu h_\nu)\right]\\
S^\mu_{(6)}&=K_F\left[H^2u^\mu + 4 TH^{\mu b} (V_b - \nu h_b)\right]\\
S^\mu_{(7)}&=K(\tilde R +2 u^a u^b \tilde R_{ab} - 3 \omega_{ab} \omega^{ab}) u^\mu - 2 K D_\nu\left(\sigma^{\mu\nu} -\frac{2}{3} P^{\mu\nu}\Theta\right)\\
&~~+ 2(D_\mu K) \left(\sigma^{\mu\nu} -\frac{2}{3} P^{\mu\nu}\Theta\right)\\
 \end{split}
 \end{equation}
 Here $D_\mu$ denotes the covariant derivative with respect to the background metric. $K_T$, $K_c$, $K_{cT}$, $K_f$, $K_{Ff}$ and $K_F$ are arbitrary functions of $T$ and $\nu$. Rest of the notations are defined below.
 \begin{equation}\label{notationintro}
 \begin{split}
& P_{\mu\nu} = u_\mu u_\nu + {\mathcal G}_{\mu\nu}~ \text{where }~{\mathcal G}_{\mu\nu} = \text{Metric}\\
&\Theta =D_\mu u^\mu,~~{\mathfrak a}_\mu = (u.D) u_\mu,~~h_\mu = \mathfrak a_\mu + P_\mu ^\alpha\left(\frac{ D_\alpha T}{T}\right)\\
&\omega_{\mu\nu} = P_\mu^\alpha P_\nu^\beta \left(\frac{D_\alpha u_\beta - D_\beta u_\alpha}{2}\right),~~\sigma_{\mu\nu} = P_\mu^\alpha P_\nu^\beta \left(\frac{D_\alpha u_\beta + D_\beta u_\alpha}{2}-\frac{\Theta}{3}{\mathcal G}_{\alpha\beta}\right)\\
&\tilde R = \text{Ricci scalar},~~\tilde R_{\mu\nu} =\text{Ricci tensor}, ~~F_{\mu\nu} = \text{Field strength},~~E_\mu = F_{\mu\nu} u^\nu\\
&V_\mu  = \frac{E_\mu}{T} - P_\mu^\alpha D_\alpha \nu,~~\bar F_{\mu\nu}=P_\mu^\alpha P_\nu^\beta F_{\alpha\beta},~~H_{\mu\nu} = \bar F_{\mu\nu} + 2 T \nu\omega_{\mu\nu}
 \end{split}
 \end{equation}
 Now we shall write the expressions for $S^\mu_{zero-divergence}$ and $S^\mu_B$
 \begin{equation}\label{finalresult4}
 \begin{split}
S^\mu_{zero-divergence} &= D_\nu\bigg[a_1(u^\mu D^\nu T - u^\nu D^\mu T)+a_2(u^\mu D^\nu \nu - u^\nu D^\mu \nu)+a_3\omega^{\mu\nu}+ a_4\bar F^{\mu\nu}\\
 &~~~~~~~~~+a_5(u^\mu V^\nu - u^\nu V^\mu)\bigg]\\
S^\mu_B &=[b_1\sigma^2 + b_2V^2 + b_3 \Theta^2 ] u^\mu + b_4 \sigma^{\mu\nu}V_\nu + b_5 \Theta V^\mu 
 \end{split}
 \end{equation}
 Here also $a_i$'s and $b_i$'s are arbitrary functions of temperature and the chemical potential and we have used the following notations.
 $$\sigma^2 = \sigma_{\mu\nu}\sigma^{\mu\nu},~~ V^2 = V_\mu V^\mu$$
 Equations \eqref{finalresult1}, \eqref{finalresult2}, \eqref{finalresult3} and \eqref{finalresult4} are the main results of this note.  
 \newline
 As we have explained before, in the course of computation we have also found the constraints on the transport coefficients for the charged fluid at second order in derivative expansion, but in a non-standard frame.  Just from symmetry analysis we could have 24 independent transport coefficients that would multiply `non-dissipative' terms (terms that do not vanish in equilibrium). These are the ones that we can constrain from the analysis of the partition function or the entropy current\cite{equipart}. Here we have determined how these 24 coefficients could be expressed in terms of the 7 parameters of the partition function. So we have implicitly found 17 constraints on the most general set of constitutive relations. These constraints are described in section(\ref{sec:stresscur})(see equations \eqref{covexpress},\eqref{nondiss1},\eqref{nondiss2},\eqref{nondiss3},\eqref{nondiss4},\eqref{nondiss5} and \eqref{diss}).

  The organization of this note is as follows. In section(\ref{sec:strategy}) we shall briefly describe the set up and the method to be used in determining the entropy current. In section(\ref{sec:partition}) we shall write the most general partition function for the charged fluid at second order in derivative expansion and shall derive the equilibrium values of the stress tensor and the charge current from it. In section(\ref{sec:stresscur}) we shall determine the most general covariant form of the stress tensor and the current that are consistent with the ones derived from the partition function.  In section(\ref{sec:entropy}) we shall construct one example of  the entropy current using the partition function. We shall also compute its divergence to explicitly show that it is non-negative upto the required order.  In section(\ref{sec:ambiguity}) we discuss the ambiguities involved in determining the entropy current and shall extend it to the most general form. Finally in section(\ref{sec:conclude}) we conclude and discuss the future directions.
  \newline
   In this note, our analysis will be restricted only to the parity even sector of the charged fluid.

\section{The method}\label{sec:strategy}
In this section we shall briefly describe the set-up and the method that we are going to use to determine the entropy current. We shall simply state the steps we need to follow without giving any justification. See \cite{entpart} for more detailed explanation.
\subsection{The basic set-up}
As mentioned in the introduction (section \ref{sec:intro}), in this note we shall study a charged fluid at second order in derivative expansion. For such a fluid system the basic variables are fluid velocity ($u^\mu$), temperature ($T$) and chemical potential ($\mu$ or $\nu =\frac{\mu}{T}$). The fluid lives on a slowly varying but otherwise arbitrary background metric denoted as ${\cal G}_{\mu\nu}$ and in presence of a background abelian gauge field whose field strength is denoted as $F_{\mu\nu}$. As usual this background electromagnetic field should also have a slow dependence on the space-time so that the whole system remains in the `fluid regime'. This means that the stress tensor and the current of this system should always admit a derivative expansion when expressed in terms of the fluid variables. We shall decompose the stress tensor and the current into an `ideal' part (that is without any derivatives) and a part involving derivative corrections. The `correction' part can be further decomposed depending on the number of space-time derivatives.
\begin{equation}\label{stresscur0}
\begin{split}
 &\text{Stress tensor}=T^{\mu\nu} = E(T,\nu) u^\mu u^\nu + P(T,\nu) P^{\mu\nu} + \pi^{\mu\nu}\\
&\text{Current}=C^\mu = Q(T,\nu) u^\mu + j^\mu\\
&\text{where}~~P^{\mu\nu} = u^\mu u^\nu + {\cal G}^{\mu\nu}
\end{split}
\end{equation}
Here $E$, $P$ and $Q$ are the energy density, pressure and the charge density respectively which are related to temperature ($T$), chemical potential ($\nu$) and entropy density ($s$) through thermodynamics.
\begin{equation}\label{thermo}
\begin{split}
&dP = s dT + Q d\mu,~~
E + P = T s + \mu Q
\end{split}
\end{equation}
$\pi^{\mu\nu}$ and $j^\mu$ contain the derivative corrections to the stress tensor and the current. They can be further decomposed as
$$\pi^{\mu\nu} = \pi_{(1)}^{\mu\nu}+ \pi_{(2)}^{\mu\nu}+\cdots,~~~~j^\mu =j^\mu_{(1)} + j^\mu_{(2)} +\cdots$$
where each term in $\pi^{\mu\nu}_{(i)}$ and $j^\mu_{(i)}$ will have exactly  $i$ space-time derivatives. Existence of an entropy current with positive divergence and the existence of a partition function impose several constraints on $\pi^{\mu\nu}$ and $j^\mu$ independently at each order. The constraints on 
$\pi_{(1)}^{\mu\nu}$ and $j^\mu_{(1)}$ have already been analysed in great detail in many places. In this note we shall analyse the next order.

However (as we have already mentioned) here our aim is not to determine the constraints on $\pi_{(2)}^{\mu\nu}$ and $j^\mu_{(2)}$, rather we would like to determine an entropy current whose divergence is non-negative on any consistent profile for the fluid variables upto third order in derivative expansion. 

\subsection{The algorithm}
Our starting assumption will be about the existence of equilibrium.
We shall assume that the fluid equations will admit at least one time-independent solution when studied on a time-independent background. We shall also assume that it is possible to generate the stress tensor and the current evaluated on this particular solution from some partition function, which is a functional of the background and its derivatives.

The algorithm has two parts. In the first part we shall use the 
 equilibrium  partition function for the system to partially fix the entropy current. In the second part we shall extend it by adding further corrections so that its divergence is positive definite upto the required order.
 
We shall write the entropy current as a sum of three terms (each terms has an independent derivative expansion).
\begin{equation}\label{subent}
\begin{split}
\text{Entropy current} =J^\mu = J^\mu_{can} + S^\mu + J^\mu_{ext}
\end{split}
\end{equation}
Here the canonical part of the entropy current is denoted as $J^\mu_{can}$. This is completely fixed in terms of the derivative corrections to the ideal part of the stress tensor and the current. The divergence of $J^\mu_{can}$ can also be computed exactly  using the fluid equations and thermodynamics.
 \begin{equation}\label{divcan}
 \begin{split}
 J^\mu_{can} &= s u^\mu - \frac{u_\nu\pi^{\mu\nu}}{T}-\nu j^\mu\\
 \nabla_\mu J^\mu_{can} &= (j^\mu u_\mu)(u.\partial\nu) -\left(\frac{u_\mu u_\nu \pi^{\mu\nu}}{T^2} \right)(u.\partial T) -\left(\frac{P_{\mu\nu}\pi^{\mu\nu}}{3T} \right)\Theta\\
 &+V_\mu j^\mu + \left(\frac{u_\nu \pi^{\mu\nu}}{T}\right)h_\mu -\left(\frac{\pi^{\mu\nu}}{T}\right)\sigma_{\mu\nu}
 \end{split}
 \end{equation}
 In equation \eqref{divcan} we have used the following standard notations and definitions for various expressions.
 \begin{equation}\label{notation0}
 \begin{split}
 &P^{\mu\nu} = {\cal G}^{\mu\nu} + u^\mu u^\nu=~\text{Projector perpendicular to $u^\mu$}\\
 &\Theta = D_\mu u^\mu,~~~\mathfrak a_\mu =(u^\nu D_\nu) u^\mu\\
 &h_\mu = (u^\nu D_\nu) u^\mu +P_\mu ^\alpha\left(\frac{ D_\alpha T }{T}\right)\\
 &V_\mu = \frac{F_{\mu\nu} u^\nu }{T}- P_\mu ^\alpha D_\alpha \nu\\
 &\sigma_{\mu\nu} = P_\mu^{\alpha} P_\nu^\beta\left[\frac{D_\alpha u_\beta +D_\beta u_\alpha }{2}- \frac{\Theta}{3} {\cal G}_{\alpha\beta}\right]\\
 &\text{Here $D_\mu$ is the covariant derivative with respect to metric ${\cal G}_{\mu\nu}$}
 \end{split}
 \end{equation}
The term $S^\mu$ will be determined using the partition function and $J^\mu_{ext}$ are the corrections that we need to add finally.

 \subsubsection{Part-1: Determining $S^\mu$}\label{strat-1}
This is the first part of the method where we shall use the equilibrium partition function. We need to characterize the most general form of the partition function at the given order we are interested in. The partition function will be a functional of the background metric and the gauge field, in a time independent situation.
So the first step would be to write down the most general time independent background metric and the gauge field.
\begin{equation}\label{equinot0}
 \begin{split}
 &\text{Metric}: ~~ds^2 =G_{\mu\nu}dx^\mu dx^\nu = - e^{2\sigma}(dt + a_i dx^i)^2 + g_{ij}dx^i dx^j\\
 &\text{Gauge~field}:~ {\cal A} = A_0 dx^0 + {\cal A}_i dx^i = A_0 dt + (A_i + a_iA_0) dx^i\\
 &\text{Inverse length of the time circle } = T_0\\
 & \text{Holonomy around time circle at } = A_0\\
 \end{split}
 \end{equation}
`$\bar\nabla_\mu$' denotes covariant derivative with respect to the full metric `$G_{\mu\nu}$' and `$\nabla_i$' denotes covariant derivative with respect to the spatial metric `$g_{ij}$'.
For the fluid variables we shall use $u^\mu,~T,~\mu$ to denote the 4-velocity, temperature and the chemical potential respectively. $u^\mu$ is normalized to $(-1)$. Instead of $\mu$ we shall often used $\nu$ as the independent variable, related to $\mu$ as $\nu = \frac{\mu}{T}$.   

Let us also fix some notations that we shall use later.
 \begin{equation}\label{notation}
 \hat u^\mu = e^{-\sigma}\{1,0,0,0\},~~\hat T = T_0 e^{-\sigma},~~\hat\nu = \frac{A_0}{T_0},~~\hat a_i = T_0 a_i
 \end{equation}
 In general if ${\cal B}(u^\mu, T,\nu)$ is some arbitrary function of fluid variables then by  $\hat{\cal B}$ we denote the same quantity evaluated on $\{\hat u^\mu,\hat T,\hat\nu\}$ and the background as given in equation \eqref{equinot0}.
 $$\hat{\cal B}={\cal B}(\hat u^\mu,\hat T,\hat\nu)$$
 We should be able to write the most general partition function as a functional of $\hat T, ~\hat\nu,~ \hat a_i.~ A_i$ and their derivatives. We shall denote the partition function as $W$.
 $$ W = \int \sqrt{g}~ L = \int\sqrt{g} \left[L_0 + L_1  + L_2 + \cdots\right]= \int\sqrt{g} \left[L_0 + L_{pert}\right]$$
  where $L_k$ is a local function of $\hat T, ~\hat\nu,~ \hat a_i.~ A_i$ with exactly $k$ space derivatives. In the last equality we have denoted the sum of all the $L_k$'s with one or higher value for $k$ as $L_{pert}$.
  
    Each $L_k$ must be a scalar under the following two diffeomorphism and the abelian gauge transformation.
  $$x^i\rightarrow x^i + y^i(\vec x),~~t\rightarrow t + c(\vec x),~~{\cal A}_i\rightarrow {\cal A}_i + \partial_i \Lambda(\vec x)$$
  These symmetries will restrict the number of possible terms at any given order.
  
  Once the partition function has been fixed upto the required order in derivative expansion we  have to perform the following operations on it to determine $S^\mu$.
 \begin{enumerate}
 \item We shall determine the variation of $W$ under small fluctuation (only upto linear order) of the background fields. We shall denote it as $\delta W$. From general principle we know \cite{equipart}, \cite{entpart} we can always rewrite $\delta W$ as
 \begin{equation}\label{deltaw}
  \delta W = \int T_0\sqrt{g} \left[-2\hat T^{\mu\nu}\delta G_{\mu\nu} + \hat C^\mu\delta{\cal A}_\mu\right] + \int\sqrt{g}\nabla_i \hat J^i
  \end{equation}
 where $\hat T^{\mu\nu}$ and $\hat C^\mu$ are the stress tensor and the current of the fluid evaluated in equilibrium. The last term is the boundary term which we usually ignore when we are interested in the constraints on the stress tensor and the current. But to determine the entropy current we have to pick up just the boundary term. 

 From equation \eqref{equinot0} it is clear that the different components of the metric fluctuations could be expressed in terms of $\delta \hat T$, $\delta\hat a_i$ and $\delta g_{ij}$ and similarly the gauge field fluctuations could be written in terms of $\delta\hat\nu$ and $\delta A_i$. By construction $\hat J^i$ will be proportional to all these fluctuation.
 \item  Now we shall introduce a very slow time dependence ( much much slower than the space variation) in all the background fields. In other words, we shall make $\hat T$, $\hat a_i$, $\hat g_{ij}$, $\hat \nu$ and $\hat A_i$ dependent on both space and time but with the constraint that 
 $$\partial_0 \mathfrak Z << \partial_i \mathfrak Z ~~\text{for any $\mathfrak Z$ that is a function of $\hat T$, $\hat a_i$, $\hat g_{ij}$, $\hat \nu$ and $\hat A_i$}$$.
 
 \item Next we shall replace all the fluctuations in $\hat J^i$ as the time derivative of the corresponding background field.
 \begin{equation}\label{replace1}
 \delta \hat T\rightarrow \partial_0\hat T,~~\delta\hat\nu\rightarrow\partial_0\hat\nu,~~\delta g^{ij}\rightarrow \partial_0 g^{ij},~~\delta \hat a_i\rightarrow \partial_0 \hat a_i,~~\delta A_i\rightarrow \partial_0 A_i
 \end{equation}
 \item Now we shall  fix a very specific current $\hat S^\mu$ whose space components are given by the boundary terms  generated in the variation of the partition function with the replacement as given in \eqref{replace1} implemented. The time component of this current is identified with $L_{pert}$.
  \begin{equation}\label{replace2} 
 \hat S^0 = e^{-\sigma}L_{pert}
 \end{equation}
 \item Finally we demand that $S^\mu$ should be such that when evaluated on $\{\hat u^\mu,\hat T,\hat\nu\}$ (with all background functions being time dependent in a manner introduced in the previous steps) it reduces to $\hat S^\mu$ upto order ${\cal O}(\partial_0)$. We should emphasize that this condition might not fix $S^\mu$ uniquely. But we shall see that using some appropriate addition of $J^\mu_{ext}$ any choice of $S^\mu$ could finally be extended to an entropy current with non-negative divergence .
  \end{enumerate}

  \subsubsection{Part-2: Determining $J^\mu_{ext}$}\label{strat-2}
  Once we have chosen a form of $S^\mu$, our goal would be to add appropriate terms to the entropy current so that full divergence could be re-expressed as sum of squares upto the required order in derivative expansion. We shall call these extra terms together as $J^\mu_{ext}$. The form of $J^\mu_{ext}$ will depend on the divergence of  $(J^\mu_{can}+ S^\mu)$. So in the second part of this method our first job would be to compute this divergence. 
    
  In equation \eqref{divcan} the divergence of $J^\mu_{can}$ has already been computed exactly and the answer is given in terms of $\pi^{\mu\nu}$ and $j^\mu$. However, to cleanly analyse the positivity of the divergence we need an expression in terms of the on-shell independent fluid data. Hence we need an explicit parametrization of the stress tensor $\pi^{\mu\nu}$ and the current $j^\mu$ in terms of the independent transport coefficients (multiplying every possible on-shell independent tensor and vector structure appearing in $\pi^{\mu\nu}$ and $j^\mu$ respectively). For this, we have to count and list the independent fluid data upto some given order in derivative expansion using the symmetry of the system. Ideally we should also fix to some fluid frame. But to keep the discussion general, we shall choose not to fix it except in strict equilibrium. Therefore our parametrization of the stress tensor and current will have some redundancy. Some of the transport coefficients will not be physical as they can be absorbed in a redefinition of velocity, temperature or chemical potential at derivative order.
  Also our parametrization should be compatible with the existence of the partition function i.e. if we evaluate $\pi^{\mu\nu}$ and $j^\mu$ in equilibrium it should reduce to what we find by varying the partition function with respect to the metric and gauge field respectively as mentioned in equation \eqref{deltaw}.

Problem of finding an appropriate parametrization for the stress tensor and the current is similar to the problem of finding an expression for $S^\mu$ since in both cases we know the answer in certain limit. So for the stress tensor and current also we could use the same replacement rules as given in \eqref{replace-set}.

Once the stress tensor and the current have been properly parametrized, we shall have an explicit expression for the divergence of the canonical entropy current in terms of the independent fluid data only. The same could be done for the explicit divergence of $S^\mu$.
Now the construction of $J^\mu_{ext}$ will depend on the expression for the divergence of $(J^\mu_{can} + S^\mu)$. It will be constructed in a way so that the  upto a given order the relevant terms in the total divergence could be written as a sum of squares. It has been explained in \cite{entpart} that it could always be done. Here we shall  explicitly see it in the example of charged fluid at second order.
The entropy current, thus constructed, will have positive definite divergence in any arbitrary fluid frame.

In the following sections we shall implement these methods to construct an entropy current for a fluid with a single abelian charge at second order in derivative expansion. As a by product we shall also get the constraints on the second order transport coefficients.

 \section{The Partition function and its variation}\label{sec:partition}
 The equilibrium values for the stress tensor and the current could be determined from the partition function and the entropy current also could be partially fixed. For this we need to take variation of the partition function with respect to the metric and the gauge field.
 
 In this section we shall construct the most general partition function for the parity even charged fluid in an arbitrary static background. We shall take its variation and would determine both the bulk and the boundary term. The bulk term would determine the equilibrium stress tensor and the current and the boundary terms are required for the construction of the entropy current.
 
The first step would be to parametrize the background in the most general form. 
 \begin{equation}\label{equinot}
 \begin{split}
 &\text{Static metric}: ~~ds^2 = - e^{2\sigma}(dt + a_i dx^i)^2 + g_{ij}dx^i dx^j\\
 &\text{Gauge~field}:~ {\cal A} = A_0 dx^0 + {\cal A}_i dx^i = A_0 dt + (A_i + a_iA_0) dx^i\\
&\text{Gauge Field strength}:~ F_{ij}= \partial_i {\cal A}_j-\partial_j {\cal A}_i,~~~\hat F_{ij}= \partial_i A_j-\partial_j A_i \\
&\text{Kaluza Klein Field strength}:~ f_{ij}= (\partial_i a_j-\partial_j a_i)\\
 &\text{Inverse lenghth of the time circle} = T_0\\
 & \text{Holonomy around time citcle} = A_0\\
 &\hat T = T_0 e^{-\sigma},~~~\hat \nu = \frac{A_0}{T_0}
 \end{split}
 \end{equation}
 
 Next we need to construct the most general  partition function at second order in derivative expansion. It should be a scalar under  space diffeomorphisms, KK gauge transformation (coordinate transformation that takes $t\rightarrow t' = t + F(\vec x),~\vec x\rightarrow\vec x' =\vec x$) and ordinary gauge transformation.
 
At second order in the parity even sector upto total derivatives we could construct the following seven scalars out of the metric and the gauge field factors that contain exactly two space-derivatives \cite{equipart}.
$$R,~(\nabla \hat T)^2,~ (\nabla\hat \nu)^2,~(\nabla\hat \nu)(\nabla \hat T),~\hat f_{ij} \hat f^{ij},~ \hat f_{ij}\hat F^{ij}, ~\hat F_{ij}\hat F^{ij}$$
where $\hat f_{ij} = T_0 f_{ij}$.
So the partition function will have seven independent parameters.  For each parameter in the partition function we shall  construct a part of $S^\mu$ that satisfies all the properties mentioned in section (\ref{strat-1}).
For convenience let us parametrize the partition function at second order in the following way.
\begin{equation}\label{partition}
\begin{split}
Z_2 &=\int \sqrt{g}\bigg[ K_T(\nabla \hat T)^2 + K_c (\nabla \hat \nu)^2 + K_{cT}(\nabla\hat \nu)(\nabla\hat T) 
+ K_f \hat f^2 + K_F \hat F^2\\
&~~~~~~~~~~~~~~ + K_{fF}(\hat F_{ij}\hat f^{ij})+ K R\bigg]
\end{split}
\end{equation}
From equation \eqref{deltaw} we see that the equilibrium stress tensor and the current could be generated by varying the partition function with respect to the background metric and the gauge field. Using chain rule of functional differentiation the fluctuations of the background could be expressed as the fluctuations in $\hat T$, $\hat\nu$, $\hat a_i$ and $A_i$. Then the explicit formula for the stress tensor and the current in terms of the partition function takes the following form.
  \begin{equation}\label{stc0}
\begin{split}
\left[\frac{\hat u^\mu \hat u^\nu T_{\mu\nu} }{\hat T^2}\right]_{equilibrium}& = \frac{1}{\sqrt{g}}\left[ \frac{\delta W}{\delta \hat T}\right]\\
\left[\hat u^\mu  C_\mu\right]_{equilibrium} &= -\frac{1}{\sqrt{g}} \left[\frac{\delta W}{\delta \hat\nu}\right]\\
\left[\hat P^i_\mu  C^\mu\right]_{equilibrium}& = \frac{T}{\sqrt{g}}\left[ \frac{\delta W}{\delta A_i}\right]\\
\left[\frac{\hat P^i_\mu \hat u_\nu  T^{\mu\nu}}{\hat T^2}\right]_{equilibrium}& = \frac{1}{\sqrt{g}}\left[\frac{\delta W}{\delta (\hat a_i)}-\nu \frac{\delta W}{\delta A_i}\right] \\
\left[\frac{\hat P_{i\alpha}\hat  P_{j\beta}  T^{\alpha\beta} }{\hat T}\right]_{equilibrium}  &= -  \frac{2 }{\sqrt{g}}\left[ \frac{\delta W}{\delta g^{ij}}\right]
\end{split}
\end{equation}
In equation \eqref{stc0} all quantities in the LHS are evaluated in equilibrium.
We have used the following notation to simplify the formula.
$$\hat u^\mu = e^{-\sigma}\{1,0,0,0\},~~\hat T = T_0 e^{-\sigma},~~\hat\nu = \frac{A_0}{T_0},~~\hat P^{\mu\nu} = \hat u^\mu \hat u^\nu + G^{\mu\nu}$$

Now we shall apply equation \eqref{stc0} to the partition function as given in equation \eqref{partition}.
\begin{equation}\label{stc1}
\begin{split}
&\frac{\hat u^\mu \hat u^\nu \pi_{\mu\nu} }{\hat T^2}|_{equilibrium}\\
 =& \left(\frac{\partial K_T}{\partial \hat T}\right)(\nabla \hat T)^2-2\nabla_i(K_T \nabla^i \hat T)+\left(\frac{\partial K_c}{\partial\hat  T}\right)(\nabla \hat \nu)^2
+\left(\frac{\partial K_{cT}}{\partial \hat T}\right)(\nabla_i \hat T)(\nabla^i\hat \nu)\\
&-\nabla_i(K_{cT} \nabla^i \hat  \nu)
+\left(\frac{\partial K_f}{\partial \hat T}\right)\left(\hat f_{ij}\hat f^{ij}\right)+\left(\frac{\partial K_{Ff}}{\partial\hat T}\right)\left(\hat f_{ij}\hat F{ij}\right) +\left(\frac{\partial K_F}{\partial \hat T}\right)\left(\hat F_{ij}\hat F{ij}\right)\\
&+\left(\frac{\partial K}{\partial \hat T}\right)R\\
\end{split}
\end{equation}
\begin{equation}\label{stc2}
\begin{split}
&\hat u^\mu j_\mu|_{equilibrium} \\
=& -\left(\frac{\partial K_T}{\partial \nu}\right)(\nabla \bar T)^2-\left(\frac{\partial K_c}{\partial \nu}\right)(\nabla \hat\nu)^2+2\nabla_i(K_c \nabla^i \hat\nu)-\left(\frac{\partial K_{cT}}{\partial \nu}\right)(\nabla_i \hat T)(\nabla^i\hat\nu)\\
&+\nabla_i(K_{cT} \nabla^i \hat T)-\left(\frac{\partial K_f}{\partial \hat \nu}\right)\left(\hat f_{ij}\hat f^{ij}\right) -\left(\frac{\partial K_F}{\partial \nu}\right)\left(\hat F_{ij}\hat F{ij}\right)-\left(\frac{\partial K_{Ff}}{\partial \hat\nu}\right)\left(\hat f_{ij}\hat F^{ij}\right)\\
&-\left(\frac{\partial K}{\partial \nu}\right)R\\
\end{split}
\end{equation}
\begin{equation}\label{stc3}
\begin{split}
&\frac{\hat P^i_\mu \hat u_\nu \pi^{\mu\nu}}{\hat T^2}|_{equilibrium}\\
=&-4\nabla_j(K_f \hat f^{ji})-4\hat \nu \nabla_j(K_F \hat F^{ij})-2\nabla_j\left(K_{Ff}\hat  F^{ji}\right) -2\hat\nu\nabla_j\left(K_{Ff} \hat \hat f^{ij}\right)\\
\end{split}
\end{equation}
\begin{equation}\label{stc4}
\begin{split}
&\frac{\hat P_{i\alpha}\hat  P_{j\beta} \pi^{\alpha\beta} }{\hat T}|_{equilibrium}\\
  =&-  2K_T \left(\nabla_i \hat T\nabla_j \hat T - \frac{g_{ij}}{2}(\nabla \hat T)^2\right) -  2K_c \left(\nabla_i \hat \nu\nabla_j \hat \nu - \frac{g_{ij}}{2}(\nabla \hat\nu)^2\right)\\
  &-  2K_{cT} \left[\left(\frac{\nabla_{i} \hat T\nabla_{j} \hat\nu+\nabla_{j} \hat T\nabla_{i} \hat\nu}{2} \right)- \frac{g_{ij}}{2}(\nabla \hat\nu)(\nabla \hat T)\right]-4K_f \left[\hat f^{ik}{{\hat f}^j}_k-\frac{g^{ij}}{4}(\hat f_{ab} \hat f^{ab})\right]\\
  &-4K_F \left[\hat  F^{ik}{\hat F^j}_k-\frac{g^{ij}}{4}(\hat F_{ab} \hat  F^{ab})\right]-4K_{Ff} \left[\left(\frac{\hat F^{ik}{\hat f^j}_k +\hat f^{ik}{\hat F^j}_k}{2}\right)-\frac{g^{ij}}{4}(\hat f_{ab} \hat F^{ab})\right]\\&
  -2K\left(R_{ij} - \frac{g_{ij}}{2}R\right) + 2\left[\nabla_i \nabla_j K - (\nabla^2 K) g_{ij}\right]
\end{split}
\end{equation}

\begin{equation}\label{stc5}
\begin{split}
\hat P^i_\mu \hat j^\mu|_{equilibrium} &=4\hat T\nabla_j(K_F \hat F^{ij})+2\hat T\nabla_j\left(K_{Ff} \hat f^{ij}\right)\\
\end{split}
\end{equation}
As mentioned before, to derive the above equations (\eqref{stc1} to \eqref{stc5}) we have ignored the total derivative pieces that are generated while taking the variation of the partition function with respect to the background. However for the entropy current it is this total derivative piece that we need to  determine.

Now we shall write this total derivative piece generated from the partition function as given in equation \eqref{partition}.
\begin{equation}\label{total}
\begin{split}
&\text{Total derivative piece}=\delta Z_2 +\int\left(2T_0 \hat\pi^{\mu\nu}\delta G_{\mu\nu} - \hat j^\mu\delta {\cal A}_\mu\right)\\
 =&\int\sqrt{g}\nabla_i\bigg[2K_T(\nabla^i \hat T)\delta\hat T +2K_c(\nabla^i \hat \nu)\delta\hat \nu+ K_{cT}\left(\delta\hat T\nabla^i \hat \nu+\delta\hat \nu\nabla^i \hat T\right)
 +4 K_f \hat f^{ij}\delta \hat a_j\\
 &~~~~~~~~~~~~~~+4 K_F\hat F^{ij} \delta A_j
 +2K_{Ff}(\hat f^{ij}\delta A_j + \hat F^{ij}\delta \hat a_j)
+K(\nabla^i\delta g^k_k - \nabla_k \delta g^{ik})\bigg]
\end{split}
\end{equation}

\section{Parametrization of the stress tensor and the current}\label{sec:stresscur}
 To parametrize the stress tensor and the current upto second order in derivative expansion we need to list all the on-shell independent scalar, vector and the tensor  structures that one can build out of one or two derivatives acting on velocity, temperature, chemical potential and the background.  Since here we are doing the most general parametrization we shall not restrict ourselves to static metric and gauge field. We shall denote the general weakly curved metric as ${\mathcal G}_{\mu\nu}$ and the field strength as $F_{\mu\nu}$. For convenience we shall further classify the independent terms into two categories, dissipative (ones that vanish in a static situation) and non-dissipative (ones that do not vanish in equilibrium). Transport coefficients that multiply the non-dissipative terms are completely fixed in terms of the coefficients appearing in the partition function. At this stage we shall assume the most general set of dissipative transport coefficients whereas for the non-dissipative part we shall simple covariantize the answer we found by varying the partition function. In other words here we shall parametrize the stress tensor and the current in such a way that when evaluated on equilibrium the non dissipative part reduces to the equilibrium values as derived from the partition function.
 
 \subsection{Classification of independent data}
 To parametrize the stress tensor and the current we need to classify all possible on-shell independent terms at first and second order in derivative expansion.
First in table[\ref{table:1storder}] we list the data with single derivative.
\begin{table}[ht]
\vspace{0.5cm}
\centering 
\begin{tabular}{|c| c| c| c|} 
\hline\hline 
Scalars &Vectors & Pseudo Vectors & Tensors \\
(1) & (3) & (2) & (1)\\ [1ex] 
\hline 
\hline
$\Theta \equiv\nabla_\mu u^\mu$ & ${\mathfrak a}_\nu = u^\mu\nabla_\mu u_\nu$ & $l^\mu \equiv \epsilon^{\mu\nu\alpha\beta}u_\nu\partial_\alpha u_\beta$ & $\sigma_{\mu\nu} \equiv\nabla_{\langle\mu} u_{\nu\rangle}$ \\
& $E^\mu \equiv F^{\mu\nu}u_\nu$ & $B^\mu \equiv \frac{1}{2}\epsilon^{\mu\nu\alpha\beta}u_\nu F_{\alpha\beta}$ &\\
& $V^\mu \equiv \left(\frac{E^\mu}{T} - P^{\mu\nu}\nabla_\nu \nu\right)$ &&\\
\hline
\hline
\end{tabular}\vspace{.5cm}
\caption{Data at 1st order in derivative}
\label{table:1storder} 
\end{table}
\noindent
Here $\Theta$, $\sigma_{\mu\nu}$ and $V^\mu$ are the dissipative terms. If we do not impose the on-shell condition, at this order we could construct two more dissipative scalars, $(u.D\nu)$ and $(u.DT)$ and one more dissipative vector $\left(h_\mu = {\mathfrak a_\mu} + P_\mu^\alpha\frac{D_\alpha T}{T}\right)$. These are the dissipative terms that appear in the replacement rule as given in  \eqref{replace-set}. But they are related to the scalars and the vectors listed in the table[\ref{table:1storder}] through the conservation equations for stress tensor and current. That is why they do not appear in the list of independent data. 

At second order the data can  be of two types, ones where both the derivatives act on a single fluid or background variable ($I_2$ type) and the others which are product of two on-shell independent first order terms (composite data).
 In table[\ref{table:I2data}] we listed $I_2$ type terms. In this list the first two scalars, first three vectors and the first two tensors are dissipative.
%

\begin{table}[ht]
\caption{$I_2$ type data from fluid variables and background} 
\vspace{0.5cm}
\centering 
\begin{tabular}{|c|c| c|} 
\hline\hline 
 Scalars &Vectors  & Tensors \\
(5) & (5) & (5)\\ [1ex] 
\hline 
\hline
$(u.D)\Theta$ &$P^{\mu\nu}D_\nu \Theta$ &$(u.D)\sigma^{\langle\mu\nu\rangle}$\\
$D_\mu V^\mu$ & $P_{\alpha\nu}D_\mu \sigma^{\mu\alpha}$ &$D^{\langle\mu}V^{\nu\rangle}$\\
$D^2\nu$& $(u.D) V^\mu$ & $D^{\langle\mu} D^{\nu\rangle}\nu$\\
$D^2 T$&$P_{\mu \alpha }D_\nu F^{\nu\alpha}$& $u_\alpha u_\beta \tilde R^{\langle\mu\alpha\nu\rangle\beta}$\\
$\tilde R$& $P^\alpha_\nu u^\mu \tilde R_{\mu\alpha}$ & $R^{\langle\mu\nu\rangle}$\\
\hline
\end{tabular}\vspace{.5cm}
\label{table:I2data} 
\end{table}

In table[\ref{table:compositedata}] we listed the composite data. There are several of them.  In the list the first 5 scalars, first 8 vectors and first 7 tensors are dissipative.

\begin{table}[ht]
\caption{Composite data from fluid variables and background} 
\vspace{0.5cm}
\centering 
\begin{tabular}{|c|c|c|} 
\hline\hline 
 Scalars &Vectors  & Tensors  \\
(11) & (12) & (13) \\ [1ex] 
\hline 
\hline
$\Theta^2$&$\Theta V^\mu$ &$ \Theta \sigma^{\mu\nu}$\\
$\sigma^2$&$\sigma^\mu_\nu V^\nu$&$ V^{\langle\mu} V^{\nu\rangle}$\\
$V^2$&$\sigma^\mu_\nu D^\nu T$&$\sigma^{\langle\mu\alpha} \sigma^{\nu\rangle}_\alpha$\\
$V.DT$&$\sigma^\mu_\nu D^\nu\nu$&$ V^{\langle\mu} D^{\nu\rangle}T$\\
$V.D\nu$&$\Theta D^\mu\nu$&$V^{\langle\mu} D^{\nu\rangle}\nu$\\
$(DT).(D\nu)$&$\Theta D^\mu T$&$\bar F^{\langle\mu\alpha} \sigma^{\nu\rangle}_\alpha$\\
$(DT)^2$&$\bar F^{\mu\nu} V_\nu$&$ \omega^{\langle\mu\alpha} \sigma^{\nu\rangle}_\alpha$\\
$(D\nu)^2$&$\omega^{\mu\nu} V_\nu$&$\omega^{\langle\mu\alpha} {\omega^{\nu\rangle}}_\alpha$\\
$F_{\mu\nu} F^{\mu\nu}$&$\bar F^{\mu\nu} D_\nu T$&$ \omega^{\langle\mu\alpha} {\bar F^{\nu\rangle}}_\alpha$\\
$F_{\mu\nu}\omega^{\mu\nu}$&$\omega^{\mu\nu} D_\nu T$&$ \bar F^{\langle\mu\alpha} {{\bar F}^{\nu\rangle}}_\alpha$\\
$\omega_{\mu\nu}\omega^{\mu\nu}$&$\bar F^{\mu\nu} D_\nu \nu$&$( D^{\langle\mu}T)( D^{\nu\rangle}T)$\\
&$\omega^{\mu\nu} D_\nu \nu$&$( D^{\langle\mu}T)( D^{\nu\rangle}\nu)$\\
&&$( D^{\langle\mu}\nu)( D^{\nu\rangle}\nu)$\\
\hline
\end{tabular}\vspace{.5cm}
\label{table:compositedata} 
\end{table}
For convenience, we briefly explain the notations used in table[\ref{table:I2data}] and table[\ref{table:compositedata}].
\begin{equation}\label{notationre}
\begin{split}
&D_\mu \equiv\text{Covariant derivative with respect to the metric $\cal G_{\mu\nu}$}\\
&P_{\mu\nu} \equiv u_\mu u_\nu + {\cal G}_{\mu\nu} = \text{Projector in the directions perpendicular to $u_\mu$}\\
&A^{\langle\mu\nu\rangle} \equiv P^{\mu\alpha} P^{\nu\beta}\left[\frac{A_{\alpha\beta} +A_{\beta\alpha}}{2} - \frac{{\cal G}_{\alpha\beta}}{3} \left(A_{\theta\phi}P^{\theta\phi}\right)\right]~~\text{for any tensor $A_{\mu\nu}$}\\
&\omega_{\mu\nu} \equiv P_\mu^\alpha P_\nu^\beta \left(\frac{D_\alpha u_\beta - D_\beta u_\alpha}{2}\right),~~\sigma_{\mu\nu} \equiv P_\mu^\alpha P_\nu^\beta \left(\frac{D_\alpha u_\beta + D_\beta u_\alpha}{2}-\frac{\Theta}{3}{\mathcal G}_{\alpha\beta}\right)\\
&\tilde R = \text{Ricci scalar},~~\tilde R_{\mu\nu} =\text{Ricci tensor},~~\tilde R_{\mu\nu\alpha\beta} = \text{Riemann tensor}\\
&F_{\mu\nu} = \text{Field strength},~~\bar F_{\mu\nu}=P_\mu^\alpha P_\nu^\beta F_{\alpha\beta},~~H_{\mu\nu} = \bar F_{\mu\nu} + 2 T \nu\omega_{\mu\nu}
\end{split}
\end{equation}
In table[\ref{table:dissipative}] we have listed only the independent dissipative terms  and giving them separate names for convenience. These are the terms which would finally lead to the local production of entropy.

\begin{table}[ht!]
\caption{The dissipative terms} 
\vspace{0.5cm}
\centering 
\begin{tabular}{|c|c|c|} 
\hline\hline 
 Scalars &Vectors  & Tensors  \\
(7) & (11) & (9) \\ [1ex] 
\hline 
\hline
${\mathfrak S}_1=(u.D)\Theta$ &${\mathcal V}^\mu_1=P^{\mu\nu}D_\nu \Theta$ &${\mathcal T}_1^{\mu\nu}=(u.D)\sigma^{\langle\mu\nu\rangle}$\\
${\mathfrak S}_2=D_\mu V^\mu$ & ${\mathcal V}^\mu_2=P_{\alpha\nu}D_\mu \sigma^{\mu\alpha}$ &${\mathcal T}_2^{\mu\nu}=D^{\langle\mu}V^{\nu\rangle}$\\
& ${\mathcal V}^\mu_3=(u.D) V^\mu$ & \\
\hline
${\mathfrak S_3}=\Theta^2$&${\mathcal V}^\mu_4=\Theta V^\mu$ &${\mathcal T}^{\mu\nu}_3 = \Theta \sigma^{\mu\nu}$\\
${\mathfrak S_4}=\sigma^2$&${\mathcal V}^\mu_5=\sigma^\mu_\nu V^\nu$&${\mathcal T}^{\mu\nu}_4 = V^{\langle\mu} V^{\nu\rangle}$\\
${\mathfrak S_5}=V^2$&${\mathcal V}^\mu_6=\sigma^\mu_\nu D^\nu T$&${\mathcal T}^{\mu\nu}_5 = V^{\langle\mu} D^{\nu\rangle}T$ \\
${\mathfrak S_6}=V.DT$&${\mathcal V}^\mu_7=\sigma^\mu_\nu D^\nu\nu$& ${\mathcal T}^{\mu\nu}_{6} = V^{\langle\mu} D^{\nu\rangle}\nu$\\
${\mathfrak S_{7}}=V.D\nu$&${\mathcal V}^\mu_{8}=\Theta D^\mu\nu$&${\mathcal T}^{\mu\nu}_7 = \sigma^{\langle\mu\alpha} \sigma^{\nu\rangle}_\alpha$ \\
&${\mathcal V}^\mu_{9}=\Theta D^\mu T$&${\mathcal T}^{\mu\nu}_{8} = \bar F^{\langle\mu\alpha} \sigma^{\nu\rangle}_\alpha$\\
&${\mathcal V}^\mu_{10}=\bar F^{\mu\nu} V_\nu$&${\mathcal T}^{\mu\nu}_{9} = \omega^{\langle\mu\alpha} \sigma^{\nu\rangle}_\alpha$\\
&${\mathcal V}^\mu_{11}=\omega^{\mu\nu} V_\nu$&\\
\hline
\end{tabular}\vspace{.5cm}
\label{table:dissipative} 
\end{table}
 
\subsection{Stress tensor and Current}
Now we shall use this classification of independent fluid data to parametrize the most general stress tensor and the current, consistent with equation \eqref{stc1}, \eqref{stc2}. \eqref{stc3}, \eqref{stc4} and \eqref{stc5}.
 From symmetry analysis we could see that the `non-ideal' part of the stress tensor and the current have to be of the following form.
\begin{equation}\label{covexpress}
\begin{split}
\pi^{\mu\nu}& = A u^\mu u^\nu + B P^{\mu\nu} + (H^\mu u^\nu + H^\nu u^\mu)+ t^{\mu\nu}\\
j^\mu &= C u^\mu  + K^\mu
\end{split}
\end{equation}
 along with the constraints
 $$u_\mu H^\mu = u_\mu K^\mu =0,~~t_{\mu\nu} u^\mu = 0$$
Here $A$, $B$, $C$, $H^\mu$, $K^\mu$ and $t^{\mu\nu}$ will be functions of the fluid variables and the background and will admit a derivative expansion starting from terms with single derivative. It turns out that this description has some redundancy as some of the terms could be absorbed in a field redefinition of velocity, temperature and the chemical potential at derivative order. Here we shall only assume that our fluid variables are defined so that they reduce to the $\{\hat u^\mu,\hat T,\hat\nu\}$ in equilibrium. This does not define the fluid variables in a non-equilibrium situation and therefore does not fix the field redefinition ambiguity completely.

However in this note our goal is  to show that we can construct an entropy current with non negative divergence whenever the stress tensor and the current are compatible with the partition function and the first order transport coefficients satisfy some inequalities. For our purpose it would be best if we could construct such a current in any arbitrary frame. So we shall choose not to fix this redefinition ambiguity and shall work with the form as given in equation \eqref{covexpress}.

As mentioned before in the expression for $\pi^{\mu\nu}$ and $j^\mu$ it is the dissipative terms that we need to multiply by arbitrary transport coefficients. The non dissipative parts are already fixed in terms of the partition function. Below we shall first write the non-dissipative parts of $A$, $B$, $C$, $H^\mu$, $K^\mu$ and $t^{\mu\nu}$ respectively.
\begin{equation}\label{nondiss1}
\begin{split}
\frac{A_{non-diss}}{T^2} =&\left(\frac{\partial K_T}{\partial T}(D_\nu T)(D^\nu T)-2D_\nu (K_T D^\nu T) +2K_T {\mathfrak a}_\mu D^\mu T \right)
+ \left[\frac{\partial K_c}{\partial T}(D_\nu \nu)(D^\nu \nu)\right]\\
&+\left[\frac{\partial K_{cT}}{\partial T}(D_\nu \nu)(D^\nu T)-D_\nu (K_{cT} D^\nu \nu) +K_{cT} {\mathfrak a}_\mu D^\mu \nu\right]\\
&+ 4T^2\omega^2 \left(\frac{\partial K_f}{\partial T}\right)
-2TH_{ab}\omega^{ab}\left(\frac{\partial K_{Ff}}{\partial T}\right)+ H^2 \left(\frac{\partial K_F}{\partial T}\right)\\
&+\left(\frac{\partial K}{\partial T}\right)  (\tilde R + 2u^\mu u^\nu \tilde R_{\mu\nu} - 3 \omega^2)
\end{split} 
\end{equation}
\begin{equation}\label{nondiss2}
\begin{split}
\frac{B_{non-diss}}{T} =&\frac{K_T}{3}P_{\mu\nu}(D^\mu T)( D^\nu T)+\frac{K_c}{3}P_{\mu\nu}(D^\mu \nu)( D^\nu \nu)+\frac{K_{cT}}{3}P_{\mu\nu}(D^\mu T)( D^\nu \nu)\\
&- \frac{4 K_f T^2}{3}\omega^2 + \frac{2 T K_{Ff}}{3}H_{ab}\omega^{ab} - \frac{ K_F}{3}H^2 \\
&+ \frac{4}{3}\left[-P^{\mu\alpha}D_\alpha D_\mu K + K\left(\frac{\tilde R - 2 u^\mu u^\nu \tilde R_{\mu\nu} +\omega^2}{4} + P^{\mu\alpha}D_\mu {\mathfrak a}_\alpha + {\mathfrak a}^2\right)\right]
\end{split}
\end{equation}
\begin{equation}\label{nondiss3}
\begin{split}
C_{non-diss} =&\left(\frac{\partial K_c}{\partial T}(D_\mu \nu)(D^\mu \nu)-2D_\mu (K_c D^\mu \nu) +2K_c {\mathfrak a}_\mu D^\mu \nu \right)
+ \left[\frac{\partial K_T}{\partial T}(D_\nu T)(D^\nu T)\right]\\
&+\left[\frac{\partial K_{cT}}{\partial \nu}(D_\mu \nu)(D^\mu T)-D_\mu (K_{cT} D^\mu \nu) +K_{cT} {\mathfrak a}_\mu D^\mu T\right]\\
&+ 4T^2\omega^2 \left(\frac{\partial K_f}{\partial \nu}\right)
-2TH_{ab}\omega^{ab}\left(\frac{\partial K_{Ff}}{\partial \nu}\right)+ H^2 \left(\frac{\partial K_F}{\partial \nu}\right)\\
&+\left(\frac{\partial K}{\partial \nu}\right)  (\tilde R + 2u^\mu u^\nu \tilde R_{\mu\nu} - 3 \omega^2)
\end{split}
\end{equation}
\begin{equation}\label{nondiss4}
\begin{split}
\frac{t^{\mu\nu}_{non-diss}}{T}=&-2K_T \left(D^{\langle \mu} T \right)\left(D^{\nu\rangle} T\right)-2K_c \left(D^{\langle \mu} \nu \right)\left(D^{\nu\rangle} \nu\right)-2K_{cT} \left(D^{\langle\mu }\nu \right)\left(D^{\nu\rangle} T\right)\\
&-16T^2 K_f \omega^{\langle\mu\alpha}{ \omega^{\nu\rangle}}_{\alpha}-4 K_F H^{\langle\mu\alpha}{ H^{\nu\rangle}}_{\alpha}-8T K_F H^{\langle\mu\alpha} {\omega^{\nu\rangle}}_{\alpha}\\
&-2\left[K \tilde R^{\langle\mu\nu\rangle} - D^{\langle\mu }D^{\nu \rangle}K + KD^{\langle\mu} {\mathfrak a}^{\nu\rangle}+
  K{\mathfrak a}^{\langle\mu} {\mathfrak a}^{\nu\rangle} - 2 K \omega^{\langle\mu\alpha}{\omega^{\nu\rangle}}_\alpha\right]
 \end{split}
 \end{equation}
\begin{equation}\label{nondiss5}
\begin{split}
\frac{H^\mu_{non-diss}}{T} =~&8 D_\nu( T^2 K_f \omega^{\mu\nu})+2 D_a(TK_{Ff}H^{a\mu})+4 \nu D_a(T^2K_{Ff}\omega^{a\mu})\\
&-\nu D_a\left[4K_FTH^{a\mu}\right]\\
K^\mu_{non-diss}=& ~4D_\nu (T^2 K_{Ff} \omega^{\nu\mu})-4D_\nu (T K_F H^{\nu\mu})\\
\end{split}
\end{equation}
By explicit evaluation we can check that the above parametrization of the non dissipative part is compatible with the partition function\footnote{For explicit evaluation we have used the material presented in section(2) of \cite{equipart}}. We should not also note that this is not unique since addition of any dissipative term to these equations (\eqref{nondiss1} to \eqref{nondiss4}) will not affect the condition that in equilibrium it reduces to what we found from the partition function. But the dissipative part is well-defined and we shall assume the most general parametrization for the dissipative part. Hence a particular choice for the non-dissipative part is allowed without any loss of generality.

Now we  shall write the dissipative part with arbitrary coefficient for each independent term.
\begin{equation}\label{diss}
\begin{split}
A&= \alpha \Theta + A_{non-diss} + \sum_{i=1}^{i=7}\alpha_i{\mathfrak S}_i,~~
B= \beta\Theta + B_{non-diss} + \sum_{i=1}^{i=7}\beta_i{\mathfrak S}_i \\
C&= \chi\Theta + C_{non-diss} + \sum_{i=1}^{i=7}\chi_i{\mathfrak S}_i \\
H^\mu&= {\mathfrak h} V^\mu + H^\mu_{non-diss} +\sum_{i=1}^{i=11}{\mathfrak h}_i{\mathcal V}^\mu_i,~~
K^\mu= \kappa V^\mu + K^\mu_{non-diss} +\sum_{i=1}^{i=11}\kappa_i{\mathcal V}^\mu_i \\
t^{\mu\nu} &= \eta\sigma^{\mu\nu} +t^{\mu\nu}_{non-diss}+\sum_{i=1}^{i=9}\tau_i{\mathcal T}^{\mu\nu}_i 
\end{split}
\end{equation}
In writing equation \eqref{diss} we have used our knowledge \cite{equipart} about the most general form of the first order stress tensor and the current in the parity even sector.

 \section{The entropy current }\label{sec:entropy}
 As explained in section \ref{strat-1} we shall decompose the entropy current as
 $$J^\mu = J^\mu_{can} + S^\mu + J^\mu_{ext}$$
 $J^\mu_{can}$ has a universal formula. $S^\mu$ could be constructed using the total derivative pieces as described in equation \eqref{total}. The expression of $J^\mu_{ext}$ will depend on the total divergence of $J^\mu_{can}$ and $S^\mu$ calculated correctly upto third order. We shall see that we can construct $J^\mu_{ext}$ in  such a way so that the total divergence can be re expressed as a sum of squares, upto a certain order.
 
  For this, we need the explicit parametrization of the stress tensor and the current in terms of transport coefficients, which we have done in section (\ref{sec:stresscur}). In this section we shall first determine $J^\mu_{can}$, $S^\mu$  and shall derive their divergence explicitly upto the appropriate order in derivative expansion. Finally we shall use this expression for divergence to construct  $J^\mu_{ext}$ .
 
 \subsection{$J^\mu_{can}$ and its divergence}
 In this subsection we shall compute the divergence of the canonical piece of the entropy current. This has already been presented in equation \eqref{divcan}. We shall repeat the same equation here for convenience.
 \begin{equation}\label{divcanrepeat}
 \begin{split}
 J^\mu_{can} &= s u^\mu - \frac{u_\nu\pi^{\mu\nu}}{T}-\nu j^\mu\\
D_\mu J^\mu_{can} &= (j^\mu u_\mu)(u.\partial\nu) -\left(\frac{u_\mu u_\nu \pi^{\mu\nu}}{T^2} \right)(u.\partial T) -\left(\frac{P_{\mu\nu}\pi^{\mu\nu}}{3T} \right)\Theta\\
 &+V_\mu j^\mu + \left(\frac{u_\nu \pi^{\mu\nu}}{T}\right)h_\mu -\left(\frac{\pi^{\mu\nu}}{T}\right)\sigma_{\mu\nu}
 \end{split}
 \end{equation}
 Now we shall substitute the decomposition for $\pi^{\mu\nu}$ and $j^\mu$ as given in equation \eqref{covexpress}.
 
 \begin{equation}\label{divcan2}
 \begin{split}
D_\mu J^\mu_{can} &= -C(u.\partial\nu) -\left(\frac{A}{T^2} \right)(u.\partial T) -\left(\frac{B}{T} \right)\Theta
 +V_\mu K^\mu -\left(\frac{H^\mu}{T}\right)h_\mu\\
 &~~~~~ -\left(\frac{t^{\mu\nu}}{T}\right)\sigma_{\mu\nu}
 \end{split}
 \end{equation}
  We should note that in equation\eqref{divcan2} the scalars $(u.\partial T)$, $(u.\partial\nu)$ and $\Theta$  and the vectors $V_\mu$ and $h_\mu$ are not independent data. They are related by the equations of motion for the fluid variables. But the equations of motion are heavily dependent on the constitutive relations which we can determine only in a derivative expansion and therefore not known upto all orders. So though the expression of the canonical entropy current is exact as it has been presented in equation\eqref{divcan2}, it will be approximate as soon as we express it in terms of independent  data. 
  
  Next we shall substitute the decomposition of the constitutive relations in dissipative and non dissipative part as we have done in equation \eqref{diss}. In the next section we shall see that the expressions that involve the non-dissipative part will manifestly cancel against the divergence of $S^\mu$ to all order in derivative expansion. For this reason, we shall not attempt to rewrite this part of the divergence in terms of independent data. We shall use the equations of motion for the rest, which involves the dissipative part of the constitutive relations and shall express it in terms of the independent data.  We shall write an expression that is accurate upto third order in derivative expansion. The expression is very messy and not all parts are important for the construction of $J^\mu_{ext}$. However we shall try to give a full expression for the divergence in steps and also mention the terms that can potentially violate the local positivity of the divergence.
  
From equation \eqref{divcanrepeat} we see that  $D_\mu J^\mu_{can}$ is a dissipative scalar i.e it vanishes in equilibrium. It will have four types of terms (see equation \eqref{divcan3}). One set could be expressed as a produce of a first order dissipative data and higher order non-dissipative data. We shall collectively denote such terms as $\Delta_{non-diss}$. These are the terms which should get cancelled against the divergence of $ S^\mu$ and so we shall not simplify such terms using any equations of motion. The second category is $\Delta_{2nd-order}$ which consists of terms that can be expressed as a product of two first order on-shell independent dissipative data. By definition, these terms are of 2nd order in derivative expansion. The other two types are denoted as $\Delta_{diss-product}$ and $\Delta_{diss-imp}$. Both of these consist of terms that are third order in derivative expansion. The terms in  $\Delta_{diss-product}$ will always have either two or three factors of first order dissipative data whereas the terms in $\Delta_{diss-imp}$ are of the form of a product of one $I_2$ type dissipative  data and one first order dissipative data.
 \begin{equation}\label{divcan3}
 \begin{split}
 &D_\mu J^\mu_{can}= \Delta_{non-diss}+\Delta_{2nd-order} + \Delta_{diss-product} +\Delta_{diss-imp} +{\cal O}(\partial^4)\\
 \end{split}
 \end{equation}
 
 \begin{equation}\label{divcan4}
 \begin{split}
\Delta_{non-diss} =~& -C_{non-diss}(u.\partial\nu) -\left(\frac{A_{non-diss}}{T^2} \right)(u.\partial T) -\left(\frac{B_{non-diss}}{T} \right)\Theta\\
 &+V_{non-diss}^\mu K_\mu - \left(\frac{H_{non-diss}^\mu}{T}\right)h_\mu
 -\left(\frac{t_{non-diss}^{\mu\nu}}{T}\right)\sigma_{\mu\nu}\\
 \Delta_{2nd-order}=~&\left[\chi({\mathfrak D}\nu )+\frac{\alpha}{T^2}({\mathfrak D}T) - \frac{\beta}{T}\right]\Theta^2 + \left(\kappa + \frac{Q{\mathfrak h}}{E + P}\right) V^2 - \frac{\eta}{T}\sigma^2\\
 \end{split}
 \end{equation}
  Where 
 \begin{equation}\label{notation1}
 \begin{split}
{\mathfrak D}M = s\left[\frac{\partial M}{\partial s}\right]_Q +  Q\left[\frac{\partial M}{\partial Q}\right]_s~\text{For any scalar $M(s,Q)$}\\
  \end{split}
 \end{equation}
 $A_{non-diss}, B_{non-diss}, C_{non-diss}, H^\mu_{non-diss}, K^\mu_{non-diss}$ and $t^{\mu\nu}_{non-diss}$ are defined in equations \eqref{nondiss1}, \eqref{nondiss2}, \eqref{nondiss3} and \eqref{nondiss4} respectively.
 See equation \eqref{diss} for the definitions of $\chi,~\alpha,~\beta,~\kappa,~{\mathfrak h}$ and $\eta$. These are roughly the transport coefficients at first order in derivative expansion in any arbitrary frame. 
 \newline
 Now we shall write the expressions for $\Delta_{diss-product}$ and $\Delta_{diss-imp}$. We shall expand te result in terms of the basis of second order independent dissipative data as listed in table[\ref{table:dissipative}].
 \begin{equation}\label{divcan5}
 \begin{split}
 \Delta_{diss-imp} = \sum_{i=1}^2s_i \Theta{\mathfrak S}_i + \sum_{i=1}^3v_i V_\mu{\mathcal V}_i^\mu + \sum_{i=1}^2 {\mathfrak t}_i\sigma_{\mu\nu} {\mathcal T}^{\mu\nu}_i  
 \end{split}
 \end{equation}
 \begin{equation}\label{divcan6}
 \begin{split}
 \Delta_{diss-product} = \sum_{i=3}^7s_i \Theta{\mathfrak S}_i+ \sum_{i=4}^9 {\mathfrak t}_i\sigma_{\mu\nu} {\mathcal T}^{\mu\nu}_i  
 \end{split}
 \end{equation}
 The coefficients ($s_i$, $v_i$ and ${\mathfrak t}_i$) are functions of temperature and the chemical potential as well as the first and second order transport coefficients ($\{\alpha,\beta,\chi,\mathfrak h,\kappa,\eta\}$ and $\{\alpha_i,\beta_i,\chi_i,\mathfrak h_i,\kappa_i,\tau_i\}$ respectively as defined in equation \eqref{diss}).
We do not need their detailed functional form to construct one example of a consistent entropy current. However, in appendix(\ref{deriv}) we have explicitly computed these coefficients using the equations of motion upto the required order.

  \subsection{$S^\mu$ and its divergence}
  In this section we shall construct the $S^\mu$ using the total derivative piece derived in equation \eqref{total}. Next we shall compute its divergence exactly without using any equation of motion. Finally we shall get  the full expression for $D_\mu(J^\mu_{can}+ S^\mu)$. We shall explicitly see how $\Delta_{non-disss}$ gets cancelled. $S^\mu$ has been constructed to ensure this cancellation. See \cite{entpart} for detailed explanation.
  \subsubsection{Construction of $S^\mu$}
   According to the algorithm described in section \ref{strat-1} first we determine what $S^\mu$ will reduce to when evaluated on $\{\hat u^\mu,\hat T,\hat\nu\}$. We shall call it as $\hat S^\mu$. Its zero component is proportional to the `derivative correction' part in the partition function and the space components are proportional to the total derivative pieces generated in the variation of the partition function (explained in equation \eqref{replace2} and the paragraph just above it). Hence the zero component and the space compoenets could be read off from equations \eqref{partition} and \eqref{total} respectively with the replacement rules as given in equation \eqref{replace1} implemented.
\begin{equation}\label{hatsmu}
\begin{split}
\hat S^0 &= e^{-\sigma}\bigg[ K_T(\nabla \hat T)^2 + K_c (\nabla \hat \nu)^2 + K_{cT}(\nabla\hat \nu)(\nabla\hat T) 
+ K_f \hat f^2 + K_F \hat F^2\\
&~~~~~~~~~~~~~~ + K_{fF}(\hat F_{ij}\hat f^{ij})+ K R\bigg]\\
\hat S^i &=-e^{-\sigma}\bigg[2K_T(\nabla^i \hat T)\partial_0\hat T +2K_c(\nabla^i \hat \nu)\partial_0\hat \nu+ K_{cT}\left(\partial_0\hat T\nabla^i \hat \nu+\partial_0\hat \nu\nabla^i \hat T\right)
 +4 K_f \hat f^{ij}\partial_0 \hat a_j\\
 &~~~~~~~~~~~~~~+4 K_F\hat F^{ij} \partial_0 A_j
 +2K_{Ff}(\hat f^{ij}\partial_0 A_j + \hat F^{ij}\partial_0 \hat a_j)
+K(\nabla^i\partial_0 g^k_k - \nabla_k \partial_0 g^{ik})\bigg]
\end{split}
\end{equation}
Now we we have to covariantize this current $\hat S^\mu$ to construct $S^\mu$. $S^\mu$ should be such that when evaluated on $\{\hat u^\mu, \hat T,\hat\nu\}$ it should reduce to $\hat S^\mu$ upto order ${\cal O}(\partial_0)^2$.
Ideally we should write down the most general expression possible for $S^\mu$ using symmetry analysis and then we should evaluate it on $\{\hat u^\mu,\hat T,\hat\nu\}$ upto order ${\cal O}(\partial_0)$ and finally equating it with $\hat S^\mu$ we should fix the undetermined coefficients in $S^\mu$. This is a method which is bound to give the correct answer for $S^\mu$ and also would cleanly characterize the part in $S^\mu$ that cannot be fixed using the partition function alone.
 
  But for our purpose it is enough to determine one possible $S^\mu$ satisfying the above condition. It turns out in most cases, once $\hat S^\mu$ is fixed, we could use some easy tricks to construct one example of $S^\mu$, without going into the detailed symmetry analysis. Basically we have to do a series of replacement to obtain a covariant expression  for $S^\mu$ from $\hat S^\mu$. The expression of $\hat S^\mu$ will contain the functions appearing in the background metric and the gauge field and their covariant space derivatives $\nabla_i$ or time derivatives $\partial_0$. However due to the diffeomorphism and gauge covariance only some specific combinations of background functions and their derivatives can appear in $\hat S^\mu$. This simplifies the replacement.
    
  Below we are listing some of the replacement rules. These are the rules that we are going to use here. However in every case we should check whether the $S^\mu$, thus constructed, is compatible with $\hat S^\mu$ by explicit evaluation.
 \begin{equation}\label{replace-set}
 \begin{split}
&\hat u^\mu\rightarrow u^\mu, ~~\hat T\rightarrow T,~~\hat\nu\rightarrow\nu\\
&\hat f_{ij}= -2\hat T \hat \omega^{ij}\rightarrow -2T\omega_{\mu\nu},~~\hat F^{\ij}\rightarrow H^{\mu\nu}\equiv \bar F^{\mu\nu}+ 2 T\nu \omega^{\mu\nu}~~\text{ where } \bar F^{\mu\nu} = P^{\mu\alpha} P^{\nu\beta} F_{\alpha\beta}\\
& e^{-\sigma}\partial_0 g^{ij}\rightarrow -2 \left(\sigma^{\mu\nu} + \frac{P^{\mu\nu}}{3}  \Theta\right),~~\frac{e^{-\sigma}}{\sqrt{g}}\partial_0\sqrt{g} \rightarrow \Theta\\
&e^{-\sigma}\partial_0 A_j = - T\left(V_\mu - \nu  h_\mu\right),
 ~~e^{-\sigma}\partial_0\hat a_j = -  T h_\mu\\
&e^{-\sigma}\partial_0\rightarrow u^\mu D_\mu,~~~\nabla_i\rightarrow  P_\mu^\nu D_\nu \\
\end{split}
 \end{equation}
 Applying this set of replacement rules as given in equations \eqref{replace-set} to the expression of $\hat S^\mu$ as given in \eqref{hatsmu} we get the following covariant form for $S^\mu$.
\begin{equation}\label{covsmu}
\begin{split}
S^\mu =&\sum_{i=1}^7 S^\mu_{(i)}\\
S^\mu_{(1)}&= K_T \left[(D_\nu T)(D^\nu T)u^\mu -2(D^\mu T)(u.\partial T)\right]\\
S^\mu_{(2)}&=K_c \left[(D_\nu \nu)(D^\nu \nu)u^\mu -2(D^\mu \nu)(u.\partial T)\right]\\
S^\mu_{(3)}&=K_{cT} \left[(D_\nu \nu)(D^\nu T)u^\mu -(D^\mu \nu)(u.\partial \nu)-(D^\mu \nu)(u.\partial T)\right]\\
S^\mu_{(4)}&=4T^2K_f\left[( \omega_{ab}\omega^{ab})u^\mu- 2\omega^{\mu\nu} h_\nu\right]\\
S^\mu_{(5)}&=K_{Ff}\left[-2T(H_{ab}~\omega^{ab})u^\mu +2T H^{\mu\nu} h_\nu-4T^2 \omega^{\mu\nu}(V_\nu -\nu h_\nu)\right]\\
S^\mu_{(6)}&=K_F\left[H^2 u^\mu + 4 TH^{\mu b} (V_b - \nu h_b)\right]\\
\end{split}
\end{equation}
\begin{equation*}
\begin{split}
S^\mu_{(7)}&=K(\tilde R + 2u^a u^b \tilde R_{ab} - 3 \omega^2) u^\mu - 2 K D_\nu\left(\sigma^{\mu\nu} -\frac{2}{3} P^{\mu\nu}\Theta\right)
+ 2(D_\mu K) \left(\sigma^{\mu\nu} -\frac{2}{3} P^{\mu\nu}\Theta\right)
\end{split}
\end{equation*}
By construction $S^\mu$ has the same number of parameters (which are arbitrary functions of $T$ and $\nu$) as that of the partition function as given in equation \eqref{partition}. We can explicitly check that when evaluated $\{\hat u^\mu,\hat T,\hat\nu\}$, the current $S^\mu$ reduces to $\hat S^\mu$ upto order ${\cal O}(\partial_0^2)$. We should note that this form of $S^\mu$ is not unique. We could always add terms to $S^\mu$ that evaluates to ${\cal O}(\partial_0^2)$ on $\{\hat u^\mu,\hat T,\hat\nu\}$. For example, $(u^\alpha  u^\beta D_\alpha D_\beta T)  u^\mu$ is one such term. It is possible to construct many other such examples. Equation \eqref{covsmu} is just one consistent choice.
\subsubsection{Divergence of $S^\mu$}
Now we have to compute the divergence. This requires a bit of algebra which we have presented in the appendices. Here we shall only quote the results.
\begin{equation}\label{divs1}
\begin{split}
D_\mu S^\mu_{(1)} =&~(u.\partial T) \left(\frac{\partial K_T}{\partial T}(D_\nu T)(D^\nu T)-2D_\nu (K_T D^\nu T) +2K_T {\mathfrak a}_\mu D^\mu T \right)\\
&+\Theta K_T\left[-(u.\partial T)^2 + \frac{1}{3}P_{\mu\nu}(D^\mu T)( D^\nu T)\right]\\
&+(u.\partial \nu) \left[\frac{\partial K_T}{\partial \nu}(D_\nu T)(D^\nu T)\right] -2K_T (D^\mu T)( D^\nu T) \sigma_{\mu\nu}
\end{split}
\end{equation}
\begin{equation}\label{divs2}
\begin{split}
D_\mu S^\mu_{(2)}
=&~(u.\partial \nu) \left(\frac{\partial K_c}{\partial \nu}(D_\nu \nu)(D^\nu \nu)-2D_\nu (K_c D^\nu \nu) +2K_c {\mathfrak a}_\mu D^\mu \nu \right)\\
&+\Theta K_c\left[-(u.\partial \nu)^2 + \frac{1}{3}P_{\mu\nu}(D^\mu \nu )(D^\nu \nu)\right]\\
&+(u.\partial T) \left[\frac{\partial K_c}{\partial T}(D_\nu \nu)(D^\nu \nu)\right] -2K_c (D^\mu \nu)( D^\nu \nu) \sigma_{\mu\nu}
\end{split}
\end{equation}
\begin{equation}\label{divs3}
\begin{split}
D_\mu S^\mu_{(3)} =&~(u.\partial \nu) \left(\frac{\partial K_{cT}}{\partial \nu}(D_\nu \nu)(D^\nu T)-D_\nu (K_{cT} D^\nu T) +K_{cT} {\mathfrak a}_\mu D^\mu T \right)\\
&+(u.\partial T) \left[\frac{\partial K_{cT}}{\partial T}(D_\nu \nu)(D^\nu T)-D_\nu (K_{cT} D^\nu \nu) +K_{cT} {\mathfrak a}_\mu D^\mu \nu\right]\\
&+\Theta K_{cT}\left[-(u.\partial \nu)(u.\partial T) + \frac{1}{3}P_{\mu\nu}(D^\mu \nu)( D^\nu T)\right]
 -2K_{cT} \left(D^{\langle\mu }\nu D^{\nu\rangle} T\right) \sigma_{\mu\nu}
\end{split}
\end{equation}
\begin{equation}\label{divs4}
\begin{split}
D_\mu S^\mu_{(4)} =&4T^2\omega^2 (u.\partial K_f) - \frac{4K_f}{3}\left(T^2\omega^2 \right)\Theta
+8h_\mu D_\nu( T^2 K_f \omega^{\mu\nu})\\
 & - 16\sigma^{\mu\nu}\left[T^2 K_f ~{\omega^\mu}_\alpha \omega^{\nu\alpha}\right]
\end{split}
\end{equation}
\begin{equation}\label{divs5}
\begin{split}
D_\mu S^\mu_{(5)} =&-2T(H_{ab}~\omega^{ab}) (u.\partial K_{Ff}) +\frac{2TK_{Ff}}{3}(H_{ab}~\omega^{ab})\Theta-8TK_{Ff}\omega^{\langle\mu a}{H_a}^{\nu\rangle}\sigma_{\mu\nu}\\
&+2 D_a(TK_{Ff}H^{ab})h_b - 4 D_a(T^2K_{Ff}\omega^{ab})(V_b - \nu h_b) + 4T^2 K_{Ff} h_a V_b \omega^{ab}\\
\end{split}
\end{equation}
\begin{equation}\label{divs6}
\begin{split}
D_\mu S^\mu_{(6)} =&4K_F\sigma^c_a H^{ab}H_{bc} -\left(\frac{K_F}{3}\right)\Theta H^2 +H^2 (u.\partial K_F)
+(V_b -\nu h_b)D_a\left[4K_FTH^{ab}\right]\\
& - 4TK_F V_b h_a H^{ab}
\end{split}
\end{equation}
In equations \eqref{divs5} and \eqref{divs6} we have used the following notation
$$H^{\mu\nu} = \bar F^{\mu\nu} + 2 T \nu \omega^{\mu\nu}$$
Finally the divergence of the last term is as follows.
\begin{equation}\label{divs7}
\begin{split}
D_\mu S^\mu_{(7)} =&~(\tilde R +2 u^\mu u^\nu \tilde R_{\mu\nu} - 3 \omega^2)(u.D K) +K\left(\sigma^2 \Theta-\frac{4}{3}\Theta^3\right)\\
&~+ \frac{4\Theta}{3}\left[-P^{\mu\alpha}D_\alpha D_\mu K + K\left(\frac{\tilde R - 2 u^\mu u^\nu \tilde R_{\mu\nu} +\omega^2}{4} + P^{\mu\alpha}D_\mu {\mathfrak a}_\alpha + {\mathfrak a}^2\right)\right]\\
&~-2\sigma_{\mu\nu}\left[K \tilde R^{\mu\nu} - D^\mu D^\nu K + KD^\mu {\mathfrak a}^\nu + K{\mathfrak a}^\mu {\mathfrak a}^\nu - 2 K \omega^{\mu\alpha}{\omega^\nu}_\alpha\right]\\
&~-4K\sigma_{\mu\nu}(u.D)\sigma^{\mu\nu} +\frac{8K}{3}\Theta (u.D)\Theta
\end{split}
\end{equation}

Adding equations \eqref{divcan3}, \eqref{divs1}, \eqref{divs2}, \eqref{divs3}, \eqref{divs4}, \eqref{divs5}, \eqref{divs6}, \eqref{divs7} we get the expression for $D_\mu(J^\mu_{can}  + S^\mu)$. Using equations \eqref{divcan4}, \eqref{divcan5} and \eqref{divcan4} we can clearly see that $\Delta_{non-diss}$ gets cancelled.
Below we are quoting the final expression.
\begin{equation}\label{jmusmudiv}
\begin{split}
&D_\mu(J_{can}^\mu + S^\mu)\\
 =~& \Delta_{2nd-order} + \Delta_{diss-imp} + \Delta_{diss-product}\\
 &+K\sigma^2 \Theta+ \left[-\frac{4K}{3} +K_T(\mathfrak D T)^2 +K_c(\mathfrak D \nu)^2  +K_{cT}(\mathfrak D \nu)(\mathfrak D T)\right]\Theta^3\\
 &-4K\sigma_{\mu\nu}(u.D)\sigma^{\mu\nu} +\frac{8K}{3}\Theta (u.D)\Theta +{\cal O}(\partial^4)\\
 \end{split}
\end{equation}
where for any function $M(s,Q)$, $\mathfrak D M$ denotes the following
$$\mathfrak D M = s\frac{\partial M}{\partial s} + Q \frac{\partial M}{\partial Q}$$
\subsection{Constructing $J^\mu_{ext}$}
In the previous subsection we have computed the divergence of $(J^\mu_{can} + S^\mu)$. 
In this subsection we shall first analyse this expression of divergence as given in equation \eqref{jmusmudiv} and we shall see that it is not manifestly positive-definite.
There are few terms that could locally change the sign of the divergence for some very special fluid profile. Next we shall construct $J^\mu_{ext}$ to cure this problem.

By construction $J^\mu_{ext}$ will be of higher order in derivative expansion. So if we want to determine the most general entropy current only upto second order in derivative expansion we could ignore $J^\mu_{ext}$. The main point is that the divergence of a second order entropy current will be of third order and whether this third order expression could be written in a positive-definite form will depend on the presence of few fourth order terms. We shall show that these necessary fourth order terms could always be generated  with arbitrary coefficients by adding an appropriate third order $J^\mu_{ext}$. See \cite{entpart},\cite{Romat},\cite{secondorder} for more elaborate explanation.

But before going to the third order analysis, we have to complete the analysis at second order i.e. we have to first find out the constraints on the first order dissipative transport coefficients. This has been worked out in detail in many places. The conditions should be such that $\Delta_{2nd-order}$ is non-negative.
For convenience, here we are quoting the expression for  $\Delta_{2nd-order}$.
$$\Delta_{2nd-order}=~\left[\chi({\mathfrak D}\nu )+\frac{\alpha}{T^2}({\mathfrak D}T) - \frac{\beta}{T}\right]\Theta^2 + \left(\kappa + \frac{Q{\mathfrak h}}{E + P}\right) V^2 - \frac{\eta}{T}\sigma^2$$
where
\begin{equation}\label{notationff}
 \begin{split}
{\mathfrak D}M = s\left[\frac{\partial M}{\partial s}\right]_Q +  Q\left[\frac{\partial M}{\partial Q}\right]_s~\text{For any scalar $M(s,Q)$}\\
  \end{split}
 \end{equation}
Now we shall treat $\sigma^2$, $\Theta^2$ and $V^2$ as three independent functions of space-time. The expression would be non-negative only if each of the three corresponding coefficients is individually non-negative. So finally the condition on the first order transport coefficients are the following.
\begin{equation}\label{condition1}
\begin{split}
\left[\chi({\mathfrak D}\nu )+\frac{\alpha}{T^2}({\mathfrak D}T) - \frac{\beta}{T}\right]>0,~~\left(\kappa + \frac{Q{\mathfrak h}}{E + P}\right) >0,~~ \frac{\eta}{T}<0
\end{split}
\end{equation}
We shall assume that each of these three coefficients are of order ${\cal O}(1)$ in magnitude (in terms of derivative expansion they are of order ${\cal O}(\partial^0)$).

 Now, by definitions the terms in $\Delta_{diss-product}$ will have at least two factors of first order dissipative data ($\Theta$, $\sigma_{\mu\nu}$ and $V^\mu$)\footnote{In the language of \cite{entpart} these are the scalars of type ${\mathfrak H}^{(many)}$.}. In the regime where the derivative expansion is valid, these terms are always suppressed compared to $\Delta_{2nd-order}$ and so they can not change the sign of the divergence. Hence these terms are not important for our purpose.  
 We should note that the terms in the third of equation \eqref{jmusmudiv} also fall in the same category and the above argument applies to them as well.
 
 Now we have to analyse the terms of the form $\Delta_{diss-imp}$. The terms here are of the form of a first order dissipative data times a second order $I_2$ type dissipative data\footnote{In the language of \cite{entpart} these are the scalars of type $\mathfrak H^{(one)}$.}. We should note that the last two terms in equation \eqref{jmusmudiv} are also of the same type as $\Delta_{diss-imp}$. These are the terms that could locally violate the positivity of the divergence since it is possible to have a fluid configuration where at a given point in space-time locally a first order dissipative data has same order of magnitude as some second order $I_2$ type data.
 \newline
But suppose at third order, the entropy  current is such that its divergence generates fourth order terms of the form $(I_2)^2$.
Then it would always be possible to absorb the $\Delta_{diss-imp}$ into  full square terms. In other words, if we consider only $\Delta_{2ndorder}$ and $\Delta_{diss-imp}$, the expression of divergence is essentially a quadratic form in the space of first order and the second order $I_2$ type dissipative data. By adding $J^\mu_{ext}$ we generate appropriate terms of the type $(I_2)^2$ so that we could finally diagonalize the quadratic form. Schematically the diagonal form will be as follows.
 \begin{equation*}
 \begin{split}
\text{Divergence}
\sim&(\text{Coefficient})_1\times(\text{1st order dissipative}+ \text{2nd order $I_2$ type dissipative})^2\\
 & +(\text{Coefficient})_2\times (\text{2nd order $I_2$ type dissipative})^2 + \text{higher order terms}
 \end{split}
 \end{equation*}
  Then we have to impose that the coefficients of these full-square terms are always positive to ensure the positivity of the divergence \cite{entpart},\cite{Romat},\cite{secondorder}. In \cite{entpart} it has been argued that it is always possible to add such higher order corrections to the entropy current so that it produces appropriate $I_2^2$ type of terms. These are the corrections which we have called $J^\mu_{ext}$. 
    Here we shall explicitly construct them for this example of 2nd order charged fluid. As mentioned before, it will be of third order in derivative expansion. However, we should emphasis that $J^\mu_{ext}$ is not  the complete or the most general construction of the third order entropy current. Its only role is to show that we do need to impose any further constraints on the second order dissipative transport coefficients to ensure a local entropy production.
  
  So our task is to construct 3rd order vectors whose divergence will have one scalar term of the form ($I_2$ type dissipative data)$^2$. Now from table[\ref{table:dissipative}] we have two $I_2$ type dissipative scalars , three vectors and two tensors. In \cite{entpart} we have a general algorithm to construct these vectors. Here we shall simply apply it for each $(I_2)^2$ type dissipative data\footnote{In the language of \cite{entpart} these are the scalars of type ${\mathfrak H}^{(zero)}$.}.
  \begin{equation}\label{53eq1}
  \begin{split}
  &{\mathfrak S_1^2} =(u.D\Theta)^2 = \frac{1}{2}D_\mu \left[u^\mu (u.D)\Theta^2\right] - \Theta D_\mu \left[u^\mu (u.D)\Theta\right]\\
  &\mathfrak S_2^2 =(D.V)^2 =D_\mu\left[V^\mu(D.V)\right]- V^\mu D_\mu(D.V)\\
 \end{split}
 \end{equation}
 \begin{equation}\label{53eq2}
 \begin{split}
  &{\mathcal V}_1^2 \sim(D^\mu\Theta)(D_\mu\Theta) = \frac{1}{2}D^2(\Theta^2)-\Theta D^2\Theta\\
 &{\mathcal V}_2^2 \sim ( D_\alpha\sigma^{\mu\alpha})(D_\beta\sigma_\mu^\beta)= D_\mu\left[\sigma^{\mu\alpha}D_\beta\sigma^\beta_\alpha\right] - \sigma^{\mu\alpha}D_\mu D_\nu \sigma^{\nu\alpha}\\
 &{\mathcal V}_3^2 \sim [(u.D )V^\mu][(u.D)V_\mu] = \frac{1}{2} D_\mu[u^\mu(u.D) V^2] - V_\mu D_\nu\left[u^\nu (u.D)V^\mu\right]\\
  \end{split}
 \end{equation}
 \begin{equation}\label{53eq3}
 \begin{split}
 &\mathcal T_1^{\mu\nu}\sim[(u.D)\sigma^{\mu\nu}][(u.D)\sigma_{\mu\nu}] = \frac{1}{2}D_\mu[u^\mu (u.D)\sigma^2]-\sigma_{\mu\nu}D_\alpha \left[u^\alpha (u.D)\sigma^{\mu\nu}\right] \\
 &\mathcal T_2^{\mu\nu} \sim( D_\mu V_\nu)( D^\mu V^\nu) + ( D_\mu V_\nu)( D_\nu V_\mu)= \frac{1}{2}D^2 V^2 - V_\nu D^2 V^\nu +D_\mu (V.D V^\mu) \\
 &~~~~~~~~~~~~~~~~~~~~~~~~~~~~~~~~~~~~~~~~~~~~~~~~~~~~~~- V_\nu D_\mu D^\nu V^\mu
  \end{split}
  \end{equation}
  In the equations of \eqref{53eq2} and \eqref{53eq3} we have used `$\sim$' sign because in these equations we have ignored the overall factors and also the projectors in the definition $\mathcal V_i^\mu$ and $\mathcal T_i^{\mu\nu}$. It is clear that the overall factors do not matter since we shall anyway have arbitrary coefficients in front of each independent term in $J^\mu_{ext}$. Ignoring the projectors might seem wrong. But the difference between a projected vector and its unprojected version is just one $(I_2)^2$ type dissipative scalar which we already know how to handle; similarly the difference between a projected tensor and an unprojected one is a $(vector)^2$ and a $(scalar)^2$, both of which are already handled in previous equations. 
  \newline
  From equations \eqref{53eq1}, \eqref{53eq2} and \eqref{53eq3} we could see that we have to choose $J^\mu_{ext}$ in the following way.
  \begin{equation}\label{jmuext1}
  \begin{split}
  J^\mu_{ext} &= \tilde s_1 u^\mu (u.D)\Theta^2 + \tilde s_2 V^\mu (D.V) +\tilde v_1 D^\mu \Theta^2  + \tilde v_2  (\sigma^{\mu\alpha}D_\nu\sigma^\nu_\alpha)  +\tilde v_3 u^\mu(u.D) V^2\\
  &+\tilde {\mathfrak t}_1 u^\mu (u.D)\sigma^2 +\tilde {\mathfrak t}_2\left[\frac{D^\mu V^2}{2} +( V.D) V^\mu\right]
  \end{split}
  \end{equation}The divergence of $J^\mu_{ext}$ will generate the required $I_2^2$ type of terms but along with that it will also generate terms of the form 
  $$(first ~ order~ dissipative~data )\times ~(3rd~order~dissipative~I_3~type ~data)$$
  where following the notation of \cite{secondorder}, $I_3$ type data denotes a third order term where all the three derivatives act on a single fluid variable.
These are again terms which can locally violate the entropy production for some very specific fluid profile and following the similar logic as mentioned above we need to add corrections to the entropy current so that $I_3^2$ type of terms are generated. Thus it might seem that we shall enter an infinite recursion loop if this is our method to construct one example of entropy current with non negative divergence. 

However here we are interested  upto a given order in derivative expansion (2nd order for the entropy current and 3rd order for the divergence). Though in $J^\mu_{ext}$ we analysed some third order pieces of the entropy current, its only purpose was to ensure that even in the rare cases where a particular first order data locally at some point in space-time is equal in magnitude to some second order data, the local entropy production is still valid.  Similarly $I_3^2$ terms would be required if we further want the divergence to be positive when some very particular first order order data is locally as small as some third order data. But since in our calculation we are anyway insensitive to corrections as small as third order in derivative expansion, we could safely truncate this procedure once we have generated $I_2^2$ type of terms.
  
  So finally the full entropy current is as follows.
  \begin{equation}\label{finalent}
  \begin{split}
  J^\mu = J^\mu_{can} + S^\mu + J^\mu_{ext} + {\cal O}(\partial^3)
  \end{split}
  \end{equation}
  where $J^\mu_{can}$, $S^\mu$ and $J^\mu_{ext}$ are defined in equations \eqref{divcanrepeat}, \eqref{covsmu} and \eqref{jmuext1} respectively. The divergence of each of the three terms in the entropy current is already calculated in the previous subsection. Combining these results we get the following final answer.
   \begin{equation}\label{finaldiverg}
  \begin{split}
 D_\mu J^\mu =& \left[\chi({\mathfrak D}\nu )+\frac{\alpha}{T^2}({\mathfrak D}T) - \frac{\beta}{T}\right]\Theta^2 + \left(\kappa + \frac{Q{\mathfrak h}}{E + P}\right) V^2 - \frac{\eta}{T}\sigma^2\\
 &+ \sum_{i=1}^2s_i' \Theta{\mathfrak S}_i + \sum_{i=1}^3v_i V_\mu{\mathcal V}_i^\mu + \sum_{i=1}^2 {\mathfrak t}'_i\sigma_{\mu\nu} {\mathcal T}^{\mu\nu}_i \\
 &+ \sum_{i=1}^ 2 \tilde s_i  \mathfrak S_i^2 +\sum_{i=1}^ 3 \tilde v_i  {\mathcal V}_i^2 + \sum_{i=1}^ 2 \tilde {\mathfrak t}_i  \mathcal T_i^2 +\Delta'_{diss-product} + {\mathcal O}(\partial^4)\\
 =& \left[\chi({\mathfrak D}\nu )+\frac{\alpha}{T^2}({\mathfrak D}T) - \frac{\beta}{T}-\sum_{i=1}^2 \frac{s_i'^2}{4\tilde s_i}\right]\Theta^2 + \left(\kappa + \frac{Q{\mathfrak h}}{E + P}-\sum_{i=1}^3 \frac{v_i^2}{4\tilde v_i}\right) V^2 \\
 &- \left[\frac{\eta}{T}+\sum_{i=1}^2 \frac{\mathfrak t_i'^2}{4\tilde {\mathfrak t}_i}\right]\sigma^2\\
  &+ \sum_{i=1}^ 2 \tilde s_i \left[ \mathfrak S_i-\frac{s_i'}{2\tilde s_i}\Theta\right]^2 +\sum_{i=1}^ 3 \tilde v_i \left[ {\mathcal V}^\mu_i-\frac{v_i}{2\tilde v_i}V^\mu\right]^2 
 + \sum_{i=1}^ 2 \tilde {\mathfrak t}'_i \left[ \mathcal T_i^{\mu\nu} -\frac{\mathfrak t_i}{2\tilde {\mathfrak t}'_i}\sigma^{\mu\nu}\right]^2\\ &+\Delta'_{diss-product} + {\mathcal O}(\partial^4)
  \end{split}
  \end{equation}
  Here we have absorbed the last two lines of equation \eqref{jmusmudiv} in the redefinition of the coefficients $s_i$ to $s_i'$, $\mathfrak t_i$ to $\mathfrak t_i'$ and $\Delta_{diss-product}$ to $\Delta_{diss-product}'$.
\begin{equation*}
\begin{split}  
  &s_i' = s_i + \frac{8K}{3} \delta_{(i,1)},~~~~~~~~{\mathfrak t}_i' = \mathfrak t_i - 4K\delta_{(i,1)}\\
  &\Delta_{diss-product}' =\Delta_{diss-product} +K\sigma^2 \Theta- \left[\frac{4K}{3} -K_T(\mathfrak D T)^2 -K_c(\mathfrak D \nu)^2  -K_{cT}(\mathfrak D \nu)(\mathfrak D T)\right]\Theta^3
  \end{split}
  \end{equation*}
  As we have argued before that $\Delta'_{diss-product}$ is not important from the point of view of positivity of the divergence. The rest will be positive definite provided
  \begin{equation}\label{cond1}
  \tilde s_i>0,~~\tilde v_i>0,~~\tilde{\mathfrak t}_i>0
  \end{equation}
  and also
  \begin{equation}\label{cond2}
  \begin{split}
  &\left[\chi({\mathfrak D}\nu )+\frac{\alpha}{T^2}({\mathfrak D}T) - \frac{\beta}{T}-\sum_{i=1}^2 \frac{{s'_i}^2}{4\tilde s_i}\right]>0\\
  &\left[\kappa + \frac{Q{\mathfrak h}}{E + P}-\sum_{i=1}^3 \frac{v_i^2}{4\tilde v_i}\right]>0\\
  &\left[\frac{\eta}{T}+\sum_{i=1}^2 \frac{{\mathfrak t'_i}^2}{4\tilde {\mathfrak t}_i}\right]<0
  \end{split}
  \end{equation}
But $\tilde s_i$, $\tilde v_i$ and $\tilde {\mathfrak t}_i$ are arbitrary coefficients  in the entropy current which we could choose to be anything. The only physical content of the above constraints \eqref{cond1} and \eqref{cond2} are the inequalities to be satisfied by the first order transport coefficients which we have already obtained in the first order analysis as given in equation \eqref{condition1}.  

Hence we see that for parity even charged fluid, if the stress tensor and the current are compatible with the existence partition function and the first order dissipative transport coefficients satisfy appropriate inequalities, then we can construct an entropy current with non-negative divergence everywhere upto second order in derivative expansion. In \cite{entpart} this has been argued abstractly to all orders. Here we explicitly see how the argument goes through for a complicated example.

 \section{The ambiguities}\label{sec:ambiguity}
  In the previous sections we have constructed one example of entropy current whose divergence would be positive definite (upto third order in derivative expansion) on any solution of the most general fluid equations. However, as we have mentioned before, this construction is not unique. In this section we shall try  to parametrize the non-uniqueness of our construction. We shall try to see what other terms we could add to the entropy current without affecting the property that its divergence is positive-definite. 
  
 Our entropy current has three parts. The first part is the canonical piece of the entropy current $J^\mu_{can}$. This is completely fixed in terms of the stress tensor and current and there is no ambiguity involved here.
 
 The next part is $S^\mu$, which has been determined from the total derivative pieces generated under a variation of the equilibrium partition function. Here we could have several other choices. Firstly the partition function is itself defined only upto total derivatives. Secondly while writing a covariant version of $S^\mu$ from $\hat S^\mu$ (see section(\ref{sec:entropy}) for notations) we could always add terms that are of order ${\cal O}(\partial_0^2)$, when evaluated on $\{\hat u^\mu,\hat T,\hat\nu\}$.  These are the ambiguities that are there within the algorithm itself as spelt out in section (\ref{sec:strategy}).
 But even after fixing these choices somehow at the level of algorithm , there is still some room for further modification of $S^\mu$. In many cases, there exist terms whose divergence vanish identically and clearly addition of such terms to the entropy current is not going to affect the condition of local entropy production.
 
 The last part of the entropy current is $J^\mu_{ext}$, but this is higher order in derivative expansion, constructed just to show that the divergence of $J^\mu_{can} + S^\mu$ could be written as a sum of squares. Our analysis is in no sense complete if we are going to consider the third order terms as well in full generality. Hence in $J^\mu_{ext}$ a lot many terms could be added, but we are not concerned about them in our present analysis.
 
 Here we shall strictly restrict ourselves to 2nd order in derivative expansion and shall try to parametrize the ambiguity in the entropy current constructed upto this order.
 
 \subsection{Addition of terms with zero divergence}
Now we shall construct the terms whose divergence vanish identically.  It is clear that any such has to be of the form $(D_\nu{\cal K}^{\mu\nu})$ where ${\cal K}^{\mu\nu}$ is an antisymmetric tensor. Since we are interested in 2nd order in derivative expansion, in our case ${\cal K}^{\mu\nu}$ has to be of first order. So here we need to count and parametrize all such antisymmetric tensors at first order in derivative expansion. Now there are two ways we  can construct them. One is by antisymmetrizing a direct product of a first order vector and $u^\mu$. From table[\ref{table:1storder}] we have three on-shell independent first order vector, $\{D_\mu T,~D_\mu \nu, ~V_\mu\}$. So this way we could construct three antisymmetric tensor. The second way is to antisymmetrize $D_\mu$ operator with some zeroth order vector ( $u_\mu$ and ${\cal A}_\mu$).  So finally there exist 5 independent constructions for ${\cal K}^{\mu\nu}$.  These are as follows.
$$\left[u^\mu D^\nu T - u^\nu D^\mu T\right],~\left[u^\mu D^\nu \nu - u^\nu D^\mu \nu\right],~\left[u^\mu V^\nu - u^\nu V^\mu\right],~\bar F^{\mu\nu},~\omega^{\mu\nu}$$
Therefore we shall have a 5 parameter ambiguity at this stage.
\begin{equation}\label{choiceA}
\begin{split}
S^\mu_{zero-divergence} =~& D_\nu\bigg[a_1(u^\mu D^\nu T - u^\nu D^\mu T)+a_2(u^\mu D^\nu \nu - u^\nu D^\mu \nu)+a_3\omega^{\mu\nu}+ a_4\bar F^{\mu\nu}\\
 &~~~~~~~~~+a_5(u^\mu V^\nu - u^\nu V^\mu)\bigg]
 \end{split}
\end{equation}

 \subsection{Ambiguity in the partition function}
 As we have mentioned that the partition function is only defined upto total derivative terms. But since it is these total derivative terms in the partition function that are required to construct the entropy current, two equivalent partition functions differing only by total derivative pieces will have two different structures for $S^\mu$. But the difference must not have any impact on the constraints imposed by condition of local entropy production. In this section we shall see that in the case of charged fluid at second order in derivative expansion the difference between two such $S^\mu$ s can always be recast in a form so that its divergence vanish identically. These are exactly the terms we have already described in the previous subsection. See \cite{equipart} for a general argument.
 
  The total derivative terms that we could have added to the partition function (as given in equation \eqref{partition}) are the following.
 \begin{equation}\label{totalderi}
 \begin{split}
 W_{total~derivative} =\int\sqrt{g}\left[\nabla_i (M_1\nabla^i\hat T) +\nabla_i (M_2\nabla^i\hat \nu) + \nabla_i (M_3 a_j \hat f^{ij}) + \nabla_i (M_4 a_j \hat F^{ij})\right]
 \end{split}
 \end{equation}
Where $M_1$, $M_2$, $M_3$ and $M_4$ are some arbitrary functions of $\hat T$ and $\hat\nu$. 
Using the same prescription as described in section (\ref{sec:strategy}) we could determine the components of ($\hat S_{total~derivative}^\mu$) from the partition function.
  \begin{equation}\label{totalcurrent}
 \begin{split}
\hat S_{total~derivative}^0 &=e^{-\sigma}\nabla_i\left[(M_1\nabla^i\hat T) + (M_2\nabla^i\hat \nu) +  (M_3 a_j \hat f^{ij}) +  (M_4 a_j \hat F^{ij})\right]\\
\hat S_{total~derivative}^i &=\partial_0\left[  M_1(\nabla^i \hat T )+M_2(\nabla^i\hat\nu)+M_3 a_j \hat f^{ij} + M_4 a_j \hat F^{ij}\right]\\
 \end{split}
 \end{equation}
We have to covariantize ($\hat S_{total~derivative}^\mu$), that is, we have find a covariant current ($S_{total~derivative}^\mu$) such that when evaluated on $\{\hat u^\mu,\hat T,\hat\nu\}$ it reduces to ($\hat S_{total~derivative}^\mu$).

Now by explicit evaluation we see that the first four terms of $S^\mu_{zero-divergence}$ reduces to $(\hat S^\mu_{total~ derivative})$. So we  could construct $(S^\mu_{total-derivative} )$ simply by replacing $a_i\rightarrow M_i,~~i=\{1,2,3,4\}$ and $a_5\rightarrow 0$.
 \begin{equation}\label{covtotal}
 \begin{split}
 S_{total~derivative}^\mu =  D_\nu\bigg[M_1(u^\mu D^\nu T - u^\nu D^\mu T)+M_2(u^\mu D^\nu \nu - u^\nu D^\mu \nu)+M_3\omega^{\mu\nu}+ M_4\bar F^{\mu\nu}\bigg]
 \end{split}
 \end{equation}

This is in accordance with the general argument presented in \cite{equipart}. We could clearly see that any total derivative term in the partition function can be absorbed as terms with zero divergence in the entropy current. Therefore the ambiguities in the partition function does not introduce any new structure in the entropy current once all the divergence free vectors are taken care of at any given order in derivative expansion.
\newline
 
 \subsection{Ambiguity in covariantizing $\hat S^\mu$}
We have determined $\hat S^\mu$ from the total derivative piece of the partition function and then we have followed some replacement rule as given in \eqref{replace-set} to determine the covariant $S^\mu$. But this is a  tricky short cut to get the covariant current. We always have the freedom to add terms to $S^\mu$ that either vanish or evaluate to terms of order ${\cal O}(\partial_0^2)$ on $\{\hat u^\mu,\hat T, \hat\nu\}$. In this subsection we would like to parametrize all such different choices that were possible at the level of covariantization of $\hat S^\mu$. 

We shall start by some counting. At any order the most general entropy current has to have the following form
$$S^\mu = (scalar)~ u^\mu + (vector)^\mu$$
Now from table[\ref{table:I2data}] and table[\ref{table:compositedata}] there are 16 scalars and 17 vectors for parity even charged fluid at second order in derivative expansion. So to begin with the most general entropy current at second order could have 33 terms. Among them 7 are already determined from the partition function and 5 more can be rewritten as terms with zero divergence (see equations \eqref{covsmu} and \eqref{choiceA}). So we still could add 21 independent terms. Among these 21, only those are allowed which evaluate to ${\cal O}(\partial_0^2)$ on $\{\hat u^\mu,\hat T, \hat\nu\}$. 
Clearly we have to look for the dissipative data in table[\ref{table:dissipative}]. It turns out that among these dissipative data only 6 satisfy the above criteria. So we have 6 choices at this stage. 
\begin{equation}\label{choiceB}
S^\mu_B = [b_1\sigma^2 + b_2V^2 + b_3 \Theta^2 ] u^\mu + b_4 \sigma^{\mu\nu}V_\nu + b_5 \Theta V^\mu +  b_6 (u.D)\Theta  u^\mu
\end{equation}
So in our case, even after using the algorithm as explained in section(\ref{sec:strategy}),  we would have an 11-parameter choice or ambiguity in determining $S^\mu$. We have already seen that the first 5 parameters $\{a_i\}$, do not have any impact on the condition of local entropy production since their divergence vanishes identically. Now we shall analyse the impact of the last 6 such parameters as given in $S^\mu_B$. We have compute the divergence of $S^\mu_B$. First we shall compute the divergence of the first 5 terms in equation \eqref{choiceB}.
\begin{equation}\label{divchoiceB}
\begin{split}
D_\mu\left[b_1\sigma^2 u^\mu\right]&= \sigma^2(u.D)b_1 + b_1\left[\sigma^2\Theta  + 2\sigma_{\mu\nu}(u.D)\sigma^{\mu\nu}\right]\\
D_\mu\left[b_2 V^2 u^\mu\right]&= V^2(u.D)b_2 + b_2\left[V^2\Theta  + 2V_{\mu}(u.D)V^\mu\right]\\
D_\mu\left[b_3 \Theta^2 u^\mu\right]&= \Theta^2(u.D)b_3 + b_3\left[\Theta^3  + 2\Theta(u.D)\Theta\right]\\
D_\mu\left[ b_4 \sigma^{\mu\nu}V_\nu\right]&= \sigma^{\mu\nu}V_\nu D_\mu b_4 + b_4\left[V_\nu D_\mu\sigma^{\mu\nu} + \sigma^{\mu\nu}D_\mu V_\nu\right]\\
D_\mu\left[ b_5 \Theta V^\mu\right]&= \Theta (V.D)b_5 +b_5\left[\Theta (D_\mu V^\mu )+ V^\mu D_\mu\Theta\right]
\end{split}
\end{equation}
From equation \eqref{divchoiceB} it is clear that addition of these five terms will simply shift the coefficients $s_i$, $v_i$ and $\mathfrak t_i$  as defined in equation \eqref{divcan5} and \eqref{divcan6}. Hence addition of these new terms to the entropy current (or rather this ambiguity in the prescription to determine the entropy current from the partition function) will not have any impact on the physical constraints on the transport coefficients.
 Now we shall analyse the sixth term in $S^\mu_B$. 
 \begin{equation}\label{divsixth}
 \begin{split}
 D_\mu[b_6 u^\mu(u.D)\Theta] &=[(u.D)\Theta][(u.D )b_6] +\frac{ b_6}{2}(u.D)\Theta^2 +b_6 (u.D)\left[(u.D)\Theta \right] 
 \end{split}
 \end{equation}
 In equation \eqref{divsixth} the last term could potentially violate the positivity of the divergence. Hence we have to set $b_6$ to zero. But we should emphasize that the algorithm we have used to determine $S^\mu$, could not fix this $b_6$ coefficient. Once we have determined $S^\mu$, we have to compute its divergence explicitly. If our choice of $S^\mu$ turns out to be such that its divergence generates a term of the form  $(u.D)\left[(u.D)\Theta \right] $ (as it appears in equation \eqref{divsixth}) we have to add a term of the form $ u^\mu(u.D)\Theta$ with appropriate coefficient to cancel it. This addition would appear as a term in $J^\mu_{ext}$. However we did not require such an addition for the particular choice of $S^\mu$ we had used in this note (see equation \eqref{covsmu} and \eqref{divs1}, \eqref{divs2}, \eqref{divs3}, \eqref{divs4}, \eqref{divs5}, \eqref{divs6} and \eqref{divs7}).
 
 So finally we have a 10 parameter-ambiguity in the final form of the entropy current, that is, if we have one example of entropy current for charged fluid at second order in derivative expansion, we could add 10 more terms to it without affecting the property that its divergence is always non-negative.

 \section{Conclusions}\label{sec:conclude}
 In this note we have constructed the entropy current for parity even charged fluid at second order in derivative expansion. We assumed that the entropy current should be such that its divergence is always positive definite on any solution of fluid equations. Secondly in equilibrium, the integration of the zero component of this current on any space-like slice should reduce to the total entropy of the system. We have used the algorithm described in \cite{entpart} to construct one example of the entropy current and then we have analysed the ambiguity that is there in the algorithm.  Finally we arrived at the most general form of the entropy current at second order in derivative expansion such that its divergence is non negative for every fluid flow consistent with the conservation equations. It has 17 free coefficients that are arbitrary functions of temperature and chemical potentials. 7 of them are generated from the equilibrium partition function and therefore impose constraints on the  transport coefficients. Rest 10 are in the form of `ambiguity' and therefore does not give any constraints on the constitutive relations.
 \newline
  In some sense these 10 coefficients are not physical in our analysis since we do not know how to measure them in any physical experiment. These are terms that are non-zero only in a time dependent solution and 5 of them contribute to the local production of entropy. In our analysis these coefficients are completely free. It is an interesting question to explore whether these coefficients also satisfy some equations among themselves or with other transport coefficients. We know that these coefficients contribute in the production of entropy in non-equilibrium flow. For any non equilibrium fluid profile that connects two particular equilibrium  we could compute the total production of entropy independently in two different ways. One is using the partition functions at the two ends and the second is integrating the entropy production over the profile. The constraint that the final answer derived using two different methods should match, might give some new non-local equations on the coefficients so far not determined from our local analysis.
\newline
As we have mentioned before that here our purpose is just to show how the algorithm presented in \cite{entpart} works for the complicated example of parity even charged fluid at second order. Also we implicitly determined the constraints on the transport coefficients to be imposed at this order. We call it implicit because we have not fixed the fluid frame to any standard one. Only restriction on our frame is that the velocity, temperature and chemical potential reduce to $\{\hat u^\mu,\hat T,\hat\nu\}$ in equilibrium. Because of this, the results presented here cannot be directly compared with the other computations of the constitutive relations, for example the holographic one done in \cite{chargedbrane},\cite{holography}. It would be a straightforward exercise to fix a fluid frame and recast the constraints on the transport coefficients in standard language. We leave that for future work. Other obvious extensions would be to complete the analysis for parity odd sectors, for multiple abelian and non-abelian charges and to other dimensions.

Finally it would be interesting to see whether or how this formalism could be extended to the case of gravity. 
For example, we know that in Einstein gravity the horizon area of a black-hole (or black-brane) plays the role of entropy and it increases in time evolution as expected from the second law of thermodynamics. When we add higher derivative corrections to gravity, the horizon area is replaced by `Wald entropy' which is known to satisfy  the first law of thermodynamics \cite{Iyer}. However we do not know whether it satisfies the second law as well except in few special cases \cite{Jacob},\cite{oz1}. Now intuitively this situation is very similar to what we study in fluid dynamics. The higher derivative corrections to Einstein equations (or in other words the $\alpha'$ corrections) are in some sense analogous to the higher derivative corrections to constitutive relations. Wald entropy is the equilibrium value of the entropy. If this formalism could be applied here, then we should expect to find an extension of Wald entropy that vanishes in equilibrium, but its time derivative is positive on every solution of the corrected equations of gravity.
\newline
The fluid-gravity duality in the context of higher derivative gravity theory along with the entropy current derived from the equilibrium partition function of the dual fluid system  also might be useful in this respect. Once we know the appropriate entropy current for the dual fluid we might attempt to pull it back to the horizon\cite{oz1},\cite{Bhatt},\cite{oz2}. This could help us constructing an out-of-equilibrium extension of Wald entropy that will satisfy the second law.

\appendix
 \section{Divergence of $S^\mu$}
Here we shall derive the equations \eqref{divs1} to \eqref{divs7}. As mentioned before, we shall not use any equation of motion for this derivation. This is essentially a rewriting for the expression of divergence in some convenient basis of off-shell independent fluid data.
For example, whenever we shall see a term of the form a term of the form $D_\mu u_\nu$, we shall decompose it in terms of $\sigma_{\mu\nu}$, $\omega_{\mu\nu}$, $\Theta$ and ${\mathfrak a}_\mu$. 
\begin{equation}\label{impdecomp}
D_\mu u_\nu = \sigma_{\mu\nu} + \omega_{\mu\nu} + \frac{\Theta}{3}P_{\mu\nu} - u_\mu {\mathfrak a}_\nu
\end{equation}
But equation \eqref{impdecomp} is an identity and true for any $u^\mu$ as long as it is normalized to $(-1)$.

\subsection{Divergence of the first term in $S^\mu$}
These are the steps required to derive equation \eqref{divs1}.
\begin{equation}\label{corrdiv1}
\begin{split}
D_\mu S^\mu_{(1)} = &~D_\mu\left[K_T\left\{ (D_\nu T)(D^\nu T)u^\mu -2(D^\mu T)(u.\partial T)\right\}\right]\\
=&~(u.\partial K_T)(D_\nu T)(D^\nu T) +\Theta K_T (D_\nu T)(D^\nu T)\\
&+ 2 K_T (D_\nu T)(u.\partial)(D^\nu T) -2D_\mu \left[K_T(D^\mu T)(u.\partial T)\right]\\
=&~(u.\partial K_T)(D_\nu T)(D^\nu T) +\Theta K_T (D_\nu T)(D^\nu T)\\
&-2D_\nu (K_T D^\nu T) (u.\partial T) -2K_T D^\mu T D^\nu T (D_\mu u_\nu)\\
=&~(u.\partial K_T)(D_\nu T)(D^\nu T) +\Theta K_T (D_\nu T)(D^\nu T)
-2D_\nu (K_T D^\nu T) (u.\partial T)\\ &-\frac{2K_T}{3} \Theta P_{\mu\nu}\left(D^\mu T D^\nu T\right) -2K_T \left(D^\mu T D^\nu T\right) \sigma_{\mu\nu} 
+2K_T {\mathfrak a}_\mu D^\mu T (u.\partial T)\\
=&~(u.\partial T) \left(\frac{\partial K_T}{\partial T}(D_\nu T)(D^\nu T)-2D_\nu (K_T D^\nu T) +2K_T {\mathfrak a}_\mu D^\mu T \right)\\
&+\Theta K_T\left[-(u.\partial T)^2 + \frac{1}{3}P_{\mu\nu}\left(D^\mu T D^\nu T\right)\right]\\
&+(u.\partial \nu) \left[\frac{\partial K_T}{\partial \nu}(D_\nu T)(D^\nu T)\right] -2K_T \left(D^\mu T D^\nu T\right) \sigma_{\mu\nu}
\end{split}
\end{equation}

\subsection{Divergence of the second term in $S^\mu$}
 These are the steps required to derive equation \eqref{divs2}.
\begin{equation}\label{corrdiv2}
\begin{split}
D_\mu S^\mu_{(2)} = &~D_\mu\left[\left\{(D_\nu \nu)(D^\nu \nu)u^\mu -2(D^\mu \nu)(u.\partial \nu)\right\}\right]\\
=&~(u.\partial K_c)(D_\nu \nu)(D^\nu \nu) +\Theta K_c (D_\alpha \nu)(D^\alpha \nu)\\
&+ 2 K_c (D_\nu \nu)(u.\partial)(D^\nu \nu) -2D_\mu \left[K_c(D^\mu \nu)(u.\partial \nu)\right]\\
=&~(u.\partial K_c)(D_\nu \nu)(D^\nu \nu) +\Theta K_c (D_\nu \nu)(D^\nu \nu)\\
&-2D_\nu (K_c D^\nu \nu) (u.\partial \nu) -2K_c D^\mu \nu D^\nu \nu (D_\mu u_\nu)\\
=&~(u.\partial K_c)(D_\nu \nu)(D^\nu \nu) +\Theta K_c (D_\nu \nu)(D^\nu \nu)
-2D_\nu (K_c D^\nu \nu) (u.\partial \nu)\\ &-\frac{2K_c}{3} \Theta P_{\mu\nu}\left(D^\mu \nu D^\nu \nu\right) -2K_c \left(D^\mu \nu D^\nu \nu\right) \sigma_{\mu\nu} +2K_c {\mathfrak a}_\mu D^\mu \nu (u.\partial \nu)\\
=&~(u.\partial \nu) \left(\frac{\partial K_c}{\partial \nu}(D_\nu \nu)(D^\nu \nu)-2D_\nu (K_c D^\nu \nu)+2K_c {\mathfrak a}_\mu D^\mu \nu \right)\\
&+\Theta K_c\left[-(u.\partial \nu)^2 + \frac{1}{3}P_{\mu\nu}\left(D^\mu \nu D^\nu \nu\right)\right]\\
&+(u.\partial T) \left[\frac{\partial K_c}{\partial T}(D_\nu \nu)(D^\nu \nu)\right] -2K_c \left(D^\mu \nu D^\nu \nu\right) \sigma_{\mu\nu}
\end{split}
\end{equation}

\subsection{Divergence of the third term in $S^\mu$} These are the steps required to derive equation \eqref{divs3}.
\begin{equation}\label{corrdiv3}
\begin{split}
D_\mu S^\mu_{(3)}=~&D_\mu\left\{K_{cT} \left[(D_\nu \nu)(D^\nu T)u^\mu -(D^\mu \nu)(u.\partial \nu)-(D^\mu \nu)(u.\partial T)\right]\right\}\\
=~&(DT)(D\nu)(u.D)K_{cT} + K_{cT} (DT)(D\nu)\Theta  + K_{cT}(u.D)\left[(DT)(D\nu)\right]\\
& - D_\mu\left[K_{cT}(D^\mu \nu)(u.\partial \nu)+K_{cT}(D^\mu \nu)(u.\partial T)\right]\\
=~&(DT)(D\nu)(u.D)K_{cT} + K_{cT} (DT)(D\nu)\Theta  + K_{cT}(D_\nu T)(u.D)\left[D^\nu \nu\right]\\
& + K_{cT}(D_\nu \nu)(u.D)\left[D^\nu T\right]
 - D_\mu\left[K_{cT}(D^\mu \nu)(u.\partial \nu)+K_{cT}(D^\mu \nu)(u.\partial T)\right]\\
=&~(u.\partial \nu) \left(\frac{\partial K_{cT}}{\partial \nu}(D_\nu \nu)(D^\nu T)-D_\nu (K_{cT} D^\nu T) +K_{cT} {\mathfrak a}_\mu D^\mu T \right)\\
&+(u.\partial T) \left[\frac{\partial K_{cT}}{\partial T}(D_\nu \nu)(D^\nu T)-D_\nu (K_{cT} D^\nu \nu) +K_{cT} {\mathfrak a}_\mu D^\mu \nu\right]\\
&+\Theta K_{cT}\left[-(u.\partial \nu)(u.\partial T) + \frac{1}{3}P_{\mu\nu}\left(D^\mu \nu D^\nu T\right)\right]
 -2K_{cT} \left(D^{\langle\mu }\nu D^{\nu\rangle} T\right) \sigma_{\mu\nu}
\end{split}
\end{equation}

\subsection{Divergence of the fourth term in $S^\mu$}
These are the steps required to derive equation \eqref{divs4}.
\begin{equation}\label{detail4a}
\begin{split}
D_\mu (T^2K_f\omega^2 u^\mu)
 =&~ T^2\omega^2 (u.\partial K_f) + K_f\left[2 T \omega^2 (u.\partial T) +T^2\omega^2\Theta+2T^2\omega^{ab}(u.D)\omega_{ab}\right]\\
\end{split}
\end{equation}
Now we shall simplify the last term in the square bracket in equation \eqref{detail4a}.
\begin{equation}\label{detail4b}
\begin{split}
&2K_fT^2\omega^{ab}(u.D)\omega_{ab}\\
=&~T^2K_f \omega_{ab}{\mathfrak f}_{cd}(u.D)\left[P^a_c P^b_d\right] + T^2K_f \omega^{ab}(u.D){\mathfrak f}_{ab}\\
=&-2T^2K_f\omega^{ab}u^cD_a{\mathfrak f}_{bc}\\
=&~2T^2K_f\omega^{ab}D_a{\mathfrak a}_b +2T^2K_f\omega^{ab}{\mathfrak f}_{bc}(D_a u^c)\\
=&~2T^2K_f\omega^{ab}D_a\left(h_b - \frac{P_b^cD_c T}{T}\right) +2T^2K_f\omega^{ab}{\mathfrak f}_{bc}(D_a u^c)\\
=&~2T^2K_f\omega^{ab}D_ah_b -2TK_f\omega^2 (u.\partial T)+4T^2K_f\omega^{ab}\omega_{bc}(D_a u^c)\\
=&~2D_a(T^2K_f\omega^{ab}h_b) -2D_a(T^2K_f\omega^{ab})h_b -2TK_f\omega^2 (u.\partial T)\\
&+4T^2K_f\omega^{ab}\omega_{bc}\left(\sigma_a^c +P^c_a \frac{\Theta}{3}\right)
\end{split}
\end{equation}
Where ${\mathfrak f}_{ab} = D_a u_b - D_b u_a$.
To go from 2nd line to 3rd line we have used the following identities.
\begin{equation}\label{identity1}
\begin{split}
&\omega_{ab}P^b_d(u.D)P^a_c =\omega_{ab} {\mathfrak a}_a{\mathfrak a}_b =0\\
&D_a \mathfrak f_{bc} + D_b \mathfrak f_{ca} + D_c \mathfrak f_{ab} =0
\end{split}
\end{equation}

Adding equation \eqref{detail4a} and \eqref{detail4b}  and multiplying both sides by an overall factor of 4, we arrive at equation \eqref{divs4}.

\subsection{Divergence of the fifth term in $S^\mu$}
These are the steps required to derive equation \eqref{divs5}.

\begin{equation}\label{detail6a}
\begin{split}
&D_\mu(-2TK_{Ff} H_{ab} \omega^{ab}u^\mu)=D_\mu(-TK_{Ff} H_{ab}~{ \mathfrak f}^{ab}u^\mu)\\
=~&-TK_{Ff}\Theta H_{ab} { \mathfrak f}^{ab} - H_{ab} { \mathfrak f}^{ab}(u.D)(K_{Ff}T)-TK_{Ff}H^{ab}(u.D){ \mathfrak f}_{ab}\\
& -TK_{Ff}[{ \mathfrak f}^{ab}(u.D)H_{ab}]\\
\end{split}
\end{equation}
Now we shall simplify the third and the fourth term separately.
\begin{equation}\label{detail6b}
\begin{split}
&-TK_{Ff} [{\mathfrak f}^{ab}(u.D)H_{ab}]\\
=~&K_{Ff}\bigg[-T{\mathfrak f}^{ab} (F_{\alpha\beta} + T\nu {\mathfrak f}_{\alpha\beta})(u.D)(P^\alpha_a P^\beta_b) - 2T \omega^{\alpha\beta}(u.D)(F_{\alpha\beta} + T\nu {\mathfrak f}_{\alpha\beta})\bigg]\\
=~&K_{Ff}\bigg[2T{\mathfrak a}_a{\mathfrak f}^{ab}E_b - 2T \omega^{\alpha\beta}(u.D)(F_{\alpha\beta} + T\nu {\mathfrak f}_{\alpha\beta})\bigg]\\
=~&K_{Ff}\bigg[4T{\mathfrak a}_a\omega^{ab}E_b- 4 T \omega^2 (u.\partial)(T\nu) -2T\omega^{\alpha\beta}(u.D) F_{\alpha\beta} - 2 T^2\nu\omega^{\alpha\beta}(u.D){\mathfrak f}_{\alpha\beta}\bigg]\\
\end{split}
\end{equation}
\begin{equation*}
\begin{split}
=~&K_{Ff}\bigg[4T{\mathfrak a}_a\omega^{ab}E_b- 4 T \omega^2 (u.\partial)(T\nu) -2T\omega^{\alpha\beta}(u.D) F_{\alpha\beta} - 2 T^2\nu\omega^{\alpha\beta}(u.D){\mathfrak f}_{\alpha\beta}\bigg]\\
=~&K_{Ff}\bigg[4T{\mathfrak a}_a\omega^{ab}E_b- 4 T \omega^2 (u.\partial)(T\nu)+4T\omega^{ab}u^c D_a F_{bc} +4 T^2\nu\omega^{ab}u^cD_a{\mathfrak f}_{bc}\bigg]\\
=~&K_{Ff}\bigg[4T{\mathfrak a}_a\omega^{ab}E_b- 4 T \omega^2 (u.\partial)(T\nu)-4T\omega^{ab}H_{bc}(D_a u^c)\\
&~~~~~~~~+4T\omega^{ab} D_a E_b -4 T^2\nu\omega^{ab}D_a{\mathfrak a}_b\bigg]\\
=~&K_{Ff}\bigg[4T^2\left(h_a -\frac{ P_a^cD_c T}{T}\right)\omega^{ab}(V_b + D_b\nu)- 4 T \omega^2 (u.T)(T\nu)-4T\omega^{ab}H_{bc}(D_a u^c)\\
&~~~~~~~~+4T\omega^{ab} D_a [T(V_b + P_b^cD_c\nu)]-4 T^2\nu\omega^{ab}D_a\left(h_b -\frac{ P_b^cD_c T}{T}\right)\bigg]\\
=~&K_{Ff}\bigg[4T^2h_a V_b\omega^{ab} -4T\omega^{ab}H_{bc}(D_a u^c) + 4T^2\omega^{ab}D_a(V_b -\nu h_b)\bigg]\\
=~&K_{Ff}\left[4T^2h_a V_b\omega^{ab} -4T\omega^{ab}H_{bc}(D_a u^c) \right]- 4D_a\left[T^2K_{Ff}\omega^{ab}\right](V_b -\nu h_b)\\
&+D_a\left[ 4T^2K_{Ff} \omega^{ab}(V_b -\nu h_b)\right]
\end{split}
\end{equation*}
In the 6th line we have used the Bianchi identities for both $F_{ab}$ and ${\mathfrak f}_{ab}$
\begin{equation}\label{identity2}
\begin{split}
&D_a F_{bc} +D_b F_{ca} +D_c F_{ab} =0\\
&D_a \mathfrak f_{bc} +D_b \mathfrak f_{ca} +D_c \mathfrak f_{ab} =0\\
\end{split}
\end{equation}
For the fourth term also we can use the similar tricks.
\begin{equation}\label{detail6c}
\begin{split}
&-TK_{Ff}[H^{ab}(u.D){\mathfrak f}_{ab}]
=2TK_{Ff}H^{ab}u^cD_a{\mathfrak f}_{bc}\\
=&~K_{Ff}\left[-2TH^{ab}D_a {\mathfrak a}_b - 4TH^{ab}\omega_{bc}(D_a u^c)\right]\\
=&~K_{Ff}\left[-2TH^{ab}D_a \left(h_b -\frac{ P_b^cD_c T}{T}\right) - 4TH^{ab}\omega_{bc}(D_a u^c)\right]\\
=&~K_{Ff}\left[ (H^{ab}{\mathfrak f}_{ab})(u.\partial T) - 4TH^{ab}\omega_{bc}(D_a u^c)\right]
+2D_a\left[TK_{Ff}H^{ab}\right] h_b \\
&-2D_a\left[TK_{Ff}H^{ab} h_b \right]\\
\end{split}
\end{equation}
Here we have used the following notations
\begin{equation}\label{notations5}
\begin{split}
&F_{\mu\nu} = D_\mu {\cal A}_\nu - D_\nu {\cal A}_\mu,~~{\mathfrak f}_{\mu\nu} = D_\mu u_\nu - D_\nu u_\mu\\
&\bar F_{\mu\nu} = P^{\mu\alpha} P^{\nu\beta} F_{\alpha\beta},~~\bar{\mathfrak f}_{\mu\nu} =2 \omega_{\mu\nu} =P^{\mu\alpha} P^{\nu\beta} {\mathfrak f}_{\alpha\beta}\\
&E_\mu = F_{\mu\nu}u^\nu,~~{\mathfrak a}_\mu = u^\nu{\mathfrak f}_{\nu\mu}\\
&H^{\mu\nu} = \bar F^{\mu\nu} + 2 T \nu \omega^{\mu\nu}
\end{split}
\end{equation}
Combining equations \eqref{detail6a}, \eqref{detail6b} and \eqref{detail6c} we could arrive at equation \eqref{divs5}.
%
\subsection{Divergence of the sixth term in $S^\mu$}
These are the steps required to derive equation \eqref{divs6}.
%

\begin{equation}\label{detail5a}
\begin{split}
 &D_\mu(K_F H^2 u^\mu)\\
=&K_F\left[H^2 \Theta + 2 H^{ab} (u.D)H_{ab}\right] + H^2 (u.\partial K_F)\\
 =&K_F\bigg[H^2 \Theta+2 H^{ab} \left[F_{pq} + T\nu{\mathfrak f}_{pq}\right](u.D)\left(P^p_a P^q_b\right) + 4\omega_{ab} H^{ab} (u.\partial)(T\nu)\\
 &+2 H^{ab}(u.D) F_{ab} +2 T \nu H^{ab}(u.D) {\mathfrak f}_{ab}\bigg] +H^2 (u.\partial K_F)\\
 =&K_F\bigg[H^2 \Theta-4 E_b{\mathfrak a}_a H^{ab} + 4\omega_{ab} H^{ab}(u.\partial)(T\nu)
 -4 H^{ab}u^cD_a F_{bc} - 4 T\nu H^{ab}u^c D_a{\mathfrak f}_{bc}\bigg]\\
 & +H^2 (u.\partial K_F)\\
 =&K_F\bigg[H^2 \Theta-4 E_b{\mathfrak a}_a H^{ab} + 4\omega_{ab} H^{ab}(u.\partial)(T\nu)
 + 4 H_{ab} H^{bc} (D_a u^c)\\
 & - 4 H^{ab}(D_a E_b) + 4 T\nu H^{ab} (D_a {\mathfrak a}_b)\bigg] +H^2 (u.\partial K_F)\\
  =&K_F\bigg\{-\frac{\Theta}{3} H^2 +4 \sigma^a_c H_{ab} H^{bc} -4(V_b +D_b\nu)\left(T h_a - D_a T\right)H^{ab}+ 4\omega_{ab} H^{ab}(u.\partial)(T\nu)\\
& -4 H^{ab} D_a\left[T(V_b + P_b^c D_c\nu)\right] + 4 T\nu H^{ab} D_a\left[h_b +\frac{P_b^cD_c T}{T}\right]\bigg\} +H^2 (u.\partial K_F)\\
=&H^2 (u.\partial K_F)-\frac{K_F\Theta}{3} H^2 +4 K_F\sigma^a_c H_{ab} H^{bc} + (V_b - \nu h_b)D_a\left[4TK_FH^{ab}\right] \\
&- 4T K_FV_b h_a H^ab-D_a \left[4TK_F(V_b - \nu h_b)H^{ab}\right]
\end{split}
\end{equation}

%
%
%

\subsection{Analysis of the seventh term in $S^\mu$}
The analysis of the seventh term in the partition function is a bit more complicated. So we are giving more details in subsection.
Taking the variation of of the partition function with respect to the metric we get the following.
\begin{equation}\label{vary7}
\begin{split}
&~\delta[K\sqrt{g} R]\\
 =&~\delta[K\sqrt{g} g^{ij}] R_{ij} + \sqrt{g}K g^{ij}\delta R_{ij}\\
 =&~\delta[K\sqrt{g} g^{ij}] R_{ij}+\sqrt{g}Kg^{ij}\left(\nabla_k \delta\Gamma^k_{ij} - \nabla_j \delta\Gamma_{ik}^k\right)\\
 =&~\delta[K\sqrt{g} g^{ij}] R_{ij}+\sqrt{g}K\nabla_i\left(g^{jk}\delta\Gamma^i_{jk} - g^{ij}\delta\Gamma^k_{jk}\right)\\
 =&~\delta[K\sqrt{g} g^{ij}] R_{ij} - \sqrt{g}\left[\delta\Gamma^i_{jk}g^{jk} - g^{ij}\delta\Gamma^k_{jk}\right](\nabla_i K)+\sqrt{g}\nabla_i\left[K(g^{jk}\delta\Gamma^i_{jk} - g^{ij}\delta\Gamma^k_{jk})\right]\\
 =&~\delta[K\sqrt{g} g^{ij}] R_{ij}-\sqrt{g}\left[\nabla^i\delta g^k_k - \nabla_j\delta g^{ij}\right](\nabla_i K)+\sqrt{g}\nabla_i\left[K(\nabla^i\delta g^k_k - \nabla_j\delta g^{ij})\right]\\
=&~\delta[K\sqrt{g} g^{ij}] R_{ij}+\sqrt{g}(\nabla^2K)\delta g^k_k-\sqrt{g}(\nabla_k\nabla_i K)\delta g^{ik} \\
&~- \sqrt{g}\nabla_i\left[(\nabla^i K)\delta g^k_k - (\nabla_k K)\delta g^{ki}\right]+\sqrt{g}\nabla_i\left[K(\nabla^i\delta g^k_k - \nabla_k \delta g^{ik})\right]
\end{split}
\end{equation}In the third line and the sixth line we have used the following formula for the variation of the Ricci tensor and the Christoffel symbols.
\begin{equation}\label{rvari}
\begin{split}
&\delta R_{ij} = \delta {R^k}_{ikj} = \nabla_k\delta\Gamma^k_{ij} -\nabla_j\delta\Gamma^k_{ik} \\
&\delta\Gamma^i_{jk} = -\frac{1}{2}\left(\nabla_j\delta g^i_k +\nabla_k\delta g^i_j -\nabla^i\delta g_{jk}\right)\\
\Rightarrow~&(g^{jk}\delta\Gamma^i_{jk} - g^{ij}\delta\Gamma^k_{jk})=\nabla^i\delta g^k_k - \nabla_j\delta g^{ij}
\end{split}
\end{equation}

From the total derivative piece we can read off the time and the space component of $\hat S^\mu$
\begin{equation}\label{hatj7}
\begin{split}
&\hat S^0_{(7)} = KRe^{-\sigma}\\
&\hat S^i_{(7)} =e^{-\sigma}\left[(\nabla^i K)\delta g^k_k - (\nabla_k K)\delta g^{ki}-K(\nabla^i\delta g^k_k - \nabla_k \delta g^{ik})\right]\\
&~~~=e^{-\sigma}\left[(\nabla_j K )\left(g^{ij}\delta g^k_k -\delta g^{ij}\right) - K \nabla_j\left(g^{ij}\delta g^k_k - \delta g^{ij}\right)\right]
\end{split}
\end{equation}
 We shall use the following substitution.
\begin{equation}\label{substi7}
\begin{split}
&\delta g^{ij}\rightarrow -2e^{\sigma}\left(\sigma^{\mu\nu} + P^{\mu\nu}\frac{\Theta}{3}\right),~~
g_{ij}\delta g^{ij}=\delta g^k_k \rightarrow -2 e^{\sigma}\Theta\\
&e^{-\sigma}\left(g^{ij}\delta g^k_k -\delta g^{ij}\right)\rightarrow \left(\sigma^{\mu\nu} -\frac{2}{3} P^{\mu\nu}\Theta\right)\\
&R\rightarrow \tilde R + u^\mu u^\nu \tilde R_{\mu\nu} - 3 \omega^2,~~\nabla_i\rightarrow D_\mu\\
\end{split}
\end{equation}
Using the substitution as given in \eqref{substi7} we shall write a covariant version for the correction to the entropy current generated by this term.
\begin{equation}\label{corr7}
\begin{split}
S^\mu_{(7)}&= K(\tilde R + 2u^a u^b \tilde R_{ab} - 3 \omega^2) u^\mu - 2 K D_\nu\left(\sigma^{\mu\nu} -\frac{2}{3} P^{\mu\nu}\Theta\right)\\
& + 2(D_\mu K) \left(\sigma^{\mu\nu} -\frac{2}{3} P^{\mu\nu}\Theta\right)
\end{split}
\end{equation}
By explicit evaluation on $\{\hat u^\mu,\hat T,\hat\nu\}$ and the background as given in equation \eqref{equinot} we could check that $S^\mu_{(7)}$ reduces to $\hat S^\mu_{(7)}$ upto order ${\cal O}(\partial_0)$.
\newline
For the convenience of computation we shall rewrite equation \eqref{corr7} in a different form  using the following identities.
 \begin{equation}\label{reqid}
\begin{split}
&P^{\alpha\nu}\tilde R_{\mu\nu} u^\mu - P^{\alpha\nu}D^\mu\sigma_{\mu\nu} +\sigma^{\mu\alpha}{\mathfrak a}_\mu - P^{\alpha\nu}D^\mu\omega_{\nu\mu} + {\mathfrak a}_\mu\omega^{\mu\alpha} + \frac{2}{3}P^{\alpha\nu}D_\nu\Theta =0\\
&KP^\mu_\nu D_\alpha\omega^{\nu\alpha} - D_\alpha(K\omega^{\mu\alpha})+( D_\alpha K) \omega^{\alpha\mu} - K\omega^2 u^\mu=0
\end{split}
\end{equation}
The steps are as follows
\begin{equation}\label{corrnew7}
\begin{split}
&~\text{Correction to the entropy current} =S_{(7)}^\mu\\
=&~K(\tilde R +2 u^a u^b \tilde R_{ab} - 3 \omega^2) u^\mu - 2 K D_\nu\left(\sigma^{\mu\nu} -\frac{2}{3} P^{\mu\nu}\Theta\right)
 + 2(D_\mu K) \left(\sigma^{\mu\nu} -\frac{2}{3} P^{\mu\nu}\Theta\right)\\
 =&~K(\tilde R +2 u^a u^b \tilde R_{ab} - 3 \omega^2) u^\mu - 2 KP^\mu_\alpha D_\nu\left(\sigma^{\alpha\nu} -\frac{2}{3} P^{\alpha\nu}\Theta\right) - 2K\left(\sigma^2 - \frac{2\Theta^2}{3}\right)u^\mu\\
& + 2(D_\mu K) \left(\sigma^{\mu\nu} -\frac{2}{3} P^{\mu\nu}\Theta\right)\\
=&~K(\tilde R +2 u^a u^b \tilde R_{ab} - 3 \omega^2) u^\mu - 2 KP^\mu_\alpha D_\nu\sigma^{\alpha\nu} +\frac{4K}{3} P^{\alpha\mu}D_\alpha\Theta +\frac{4K}{3}\Theta D_\nu(P^{\mu\nu})\\
& - 2K\left(\sigma^2 - \frac{2\Theta^2}{3}\right)u^\mu
+ 2(D_\mu K) \left(\sigma^{\mu\nu} -\frac{2}{3} P^{\mu\nu}\Theta\right)\\
=&~2K\left(\frac{\tilde R}{2}{\mathcal G}^{\mu\nu}-\tilde R^{\mu\nu}\right)u_\nu - 3 K\omega^2 u^\mu + 2 K P^\mu_\nu D_\alpha \omega^{\nu\alpha}-2K{\mathfrak a}_\alpha \omega^{\alpha\mu}\\
&~+ 2 (D_\nu K - K{\mathfrak a}_\nu)\left(\sigma^{\mu\nu} -\frac{2 P^{\mu\nu}}{3}\Theta\right) - 2K\left(\sigma^2 - \frac{2\Theta^2}{3}\right)u^\mu\\
=&~ 2K \left(\frac{\tilde R}{2}{\mathcal G}^{\mu\nu}-\tilde R^{\mu\nu}\right)u_\nu - K \omega^2 u^\mu +2\left(\nabla_\nu K - K{\mathfrak a}_\nu\right)\left(\sigma^{\mu\nu} -\omega^{\mu\nu} + \frac{2\Theta}{3} P^{\mu\nu}\right) \\
&- 2K\left(\sigma^2 - \frac{2\Theta^2}{3}\right)u^\mu\\
\end{split}
\end{equation}
From third line to fourth line we have used the first identity and in the final step we have used the second identity as given in  equation \eqref{reqid}.
Now we shall compute the the divergence of the above four terms separately.
\newline
{\it{The divergence of the first term}}:
\begin{equation}\label{detail7a}
\begin{split}
&~D_\mu\left[2K \left(\frac{\tilde R}{2}{\mathcal G}^{\mu\nu}-\tilde R^{\mu\nu}\right)u_\nu\right]\\
=&~ K \tilde R \Theta  + \tilde R (u.D K) - 2 R^{\mu\nu}D_\mu(K u_\nu)\\
=&~ K \tilde R \Theta  + (\tilde R + 2 \tilde R_{ab} u^a u^b)(u.D K) - 2 P^\mu_{\alpha}\tilde R^{\alpha\nu} u_\nu D_\mu K - 2K \tilde R^{\mu\nu}\left(\sigma_{\mu\nu} + P_{\mu\nu} \frac{\Theta}{3}- u_\mu {\mathfrak a}_\nu\right)\\
=&~ \frac{K}{3} (\tilde R- 2 \tilde R_{ab} u^a u^b) \Theta  + (\tilde R + 2 \tilde R_{ab} u^a u^b)(u.D K) - 2 K \tilde R^{\mu\nu}\sigma_{\mu\nu}\\
&~- 2 P^\mu_\alpha \tilde R^{\alpha\nu}u_\nu(D_\mu K - K{\mathfrak a}_\mu)
\end{split}
\end{equation}
{\it{The divergence of the second term}}:
\begin{equation}\label{detail7b}
\begin{split}
&~D_\mu(-K\omega^2 u^\mu)\\
=&~-\omega^2 (u.D K)-K\omega^2\Theta - 2 K \omega^{ab}(u.D)\omega_{ab}\\
=&~-\omega^2 (u.D K)-K\omega^2\Theta - 2 K \omega^{ab}u^\mu D_a D_\mu u_b-2K \omega^{ab}u^\mu[d_\mu,d_a]u_b\\
=&~-\omega^2 (u.D K)-K\omega^2\Theta -2K\omega^{ab}D_a{\mathfrak a}_b + 2 K \omega^{ab}(D_a u^\mu)(D_\mu u_b)\\
=&~-\omega^2 (u.D K)+\frac{K}{3}\omega^2\Theta-2K\omega^{ab}D_a{\mathfrak a}_b + 4 K \omega^{\mu\alpha}\omega_{\nu\alpha}\sigma^\nu_\mu
\end{split}
\end{equation}
{\it{The divergence of the third term}}:
\begin{equation}\label{detail7c}
\begin{split}
&~2D_\mu\bigg[\left(D_\nu K - K{\mathfrak a}_\nu\right)\left(\sigma^{\mu\nu} -\omega^{\mu\nu} - \frac{2\Theta}{3} P^{\mu\nu}\right)\bigg]\\
=&~2D_\mu\left[\left(D_\nu K - K{\mathfrak a}_\nu\right)\right]\left(\sigma^{\mu\nu} -\omega^{\mu\nu} - \frac{2\Theta}{3} P^{\mu\nu}\right)\\
&~ +2\left(D_\nu K - K{\mathfrak a}_\nu\right)D_\mu\left[\left(\sigma^{\mu\nu} -\omega^{\mu\nu} - \frac{2\Theta}{3} P^{\mu\nu}\right)\right]\\
=&~2D_\mu\left[\left(D_\nu K - K{\mathfrak a}_\nu\right)\right]\left(\sigma^{\mu\nu} -\omega^{\mu\nu} - \frac{2\Theta}{3} P^{\mu\nu}\right)\\
&~+2\left(D_\alpha K - K{\mathfrak a}_\alpha\right)\left(P^\alpha_\nu D_\mu\sigma^{\nu\mu} +P^\alpha_\nu D_\mu\omega^{\nu\mu} - \frac{2}{3} P^{\alpha\mu}D_\mu\Theta\right)\\
&~+ 2 (u.D K) \sigma^2 - 2 (u.D K) \omega^2 - \frac{4}{3}(u.D K) \Theta^2 - \frac{4}{3}({\mathfrak a}.DK)\Theta + \frac{4K}{3} {\mathfrak a}^2\Theta\\
=&~2D_\mu\left[\left(D_\nu K - K{\mathfrak a}_\nu\right)\right]\left(\sigma^{\mu\nu} -\omega^{\mu\nu} - \frac{2\Theta}{3} P^{\mu\nu}\right)
+2\left(D_\alpha K - K{\mathfrak a}_\alpha\right)\left[P^\alpha_\nu \tilde R_{\mu}^{\nu}u^\mu + {\mathfrak a}_\mu(\sigma^{\mu\alpha} + \omega^{\mu\alpha})\right]\\
&~+ 2 (u.D K) \sigma^2 - 2 (u.D K) \omega^2 - \frac{4}{3}(u.D K) \Theta^2 - \frac{4}{3}({\mathfrak a}.DK)\Theta + \frac{4K}{3} {\mathfrak a}^2\Theta\\
\end{split}
\end{equation}
In the last line we have again used the first identity in equation \eqref{reqid}.
\newline
{\it{The divergence of the fourth term}}:
\begin{equation}\label{detail7d}
\begin{split}
&D_\mu\left[- 2K\left(\sigma^2 - \frac{2\Theta^2}{3}\right)u^\mu\right]\\
 =& -2\left(\sigma^2 - \frac{2\Theta^2}{3}\right)(u.DK + K\Theta) -4K\sigma_{\mu\nu}(u.D)\sigma^{\mu\nu} +\frac{8K}{3}\Theta (u.D)\Theta
\end{split}
\end{equation}
Adding equations \eqref{detail7a}, \eqref{detail7b}, \eqref{detail7c} and \eqref{detail7d} we arrive at the expression of divergence as given in equation \eqref{divs7}.

\section{Derivation for the coefficients $s_i$, $v_i$ and ${\mathfrak t_i}$}\label{deriv}
 In this section we shall present the explicit expressions for the coefficients that appear in the divergence of $J^\mu_{can}$ (see equation \eqref{divcan5} and \eqref{divcan6} for the definition of these coefficients).
 \begin{equation}\label{app:eq1}
 \begin{split}
 s_1=~& -\left[\frac{T^2\chi_1 ({\mathfrak D}\nu) + \alpha_1({\mathfrak D}T) + T\beta_1}{T^2}\right]
 -\left[\chi\left(\frac{\partial\nu}{\partial s}\right)+\frac{\alpha}{T^2}\left(\frac{\partial T}{\partial s}\right)\right]\left(\nu\chi - \frac{\alpha}{T}\right)\\
 &+\chi\left[\chi\left(\frac{\partial\nu}{\partial Q}\right)
 +\frac{\alpha}{T^2}\left(\frac{\partial T}{\partial Q}\right)\right]\\
  s_2=~&-\left[\frac{T^2\chi_2 ({\mathfrak D}\nu) + \alpha_2({\mathfrak D}T) + T\beta_2}{T^2}\right] -\left[\chi\left(\frac{\partial\nu}{\partial s}\right)+\frac{\alpha}{T^2}\left(\frac{\partial T}{\partial s}\right)\right]\left(\nu\kappa - \frac{\mathfrak h}{T}\right)\\
 &+\kappa\left[\chi\left(\frac{\partial\nu}{\partial Q}\right)
 +\frac{\alpha}{T^2}\left(\frac{\partial T}{\partial Q}\right)\right]\\
 \end{split}
 \end{equation}
 
\begin{equation}\label{app:eq2}
 \begin{split}
 &v_1= \left(\kappa_1 - \frac{Q\mathfrak h_1}{E +P}\right) + \frac{\mathfrak h \beta}{E + P},~~v_2= \left(\kappa_3 - \frac{Q\mathfrak h_3}{E +P}\right) + \frac{\mathfrak h \eta }{E + P}
 \\
  &v_3=\left(\kappa_2 - \frac{Q\mathfrak h_2}{E +P}\right) +\frac{\mathfrak h^2 }{E + P}~~~~~
  {\mathfrak t}_1 =- \frac{\tau_1}{T},~~~~~~ {\mathfrak t}_2 =- \frac{\tau_2}{T}
 \end{split}
 \end{equation}
 
\begin{equation}\label{app:eq3}
\begin{split}
s_3 =~& -\left[\frac{T^2\chi_3 ({\mathfrak D}\nu) + \alpha_3({\mathfrak D}T) + T\beta_3}{T^2}\right]+\left[\chi\left(\frac{\partial\nu}{\partial Q}\right)
 +\frac{\alpha}{T^2}\left(\frac{\partial T}{\partial Q}\right)\right](\chi-\mathfrak D\chi)\\
 &-\left[\chi\left(\frac{\partial\nu}{\partial s}\right)+\frac{\alpha}{T^2}\left(\frac{\partial T}{\partial s}\right)\right]\left[\nu(\chi-\mathfrak D \chi) - \frac{(\alpha -\mathfrak D \alpha+\beta)}{T}\right]\\
 s_4 =~&\left(\frac{\partial\nu}{\partial s}\right)\frac{\chi\eta}{T} +\left(\frac{\partial T}{\partial s}\right)\frac{\alpha\eta}{T^3}
-\left[\frac{T^2\chi_4({\mathfrak D}\nu) + \alpha_4({\mathfrak D}T) + T\beta_4}{T^2}\right] - \frac{\tau_3}{T}\\
s_5 =~&-\left[\chi\left(\frac{\partial\nu}{\partial s}\right) +\frac{\alpha}{T^2}\left(\frac{\partial T}{\partial s}\right)\right]\left(\kappa - \frac{Q\mathfrak h}{E +P}\right)
-\left[\frac{T^2\chi_5({\mathfrak D}\nu) + \alpha_5({\mathfrak D}T) + T\beta_5}{T^2}\right]\\
&+\left(\kappa_4 - \frac{Q\mathfrak h_4}{E +P}\right) -\left( \frac{\mathfrak h T }{E + P}\right)\left[ \chi  - \frac{Q (\alpha + \beta)}{E + P} + \frac{3{\mathfrak D}{\mathfrak h}-{4\mathfrak h}}{3T} \right]\\
\end{split}
\end{equation}
\begin{equation*}
\begin{split}
s_6 =~-& \left[\chi\left(\frac{\partial\nu}{\partial s}\right) +\frac{\alpha}{T^2}\left(\frac{\partial T}{\partial s}\right)\right]\frac{\partial}{\partial T}\left(\nu \mathfrak h-\kappa T\right)
-\left[\frac{T^2\chi_6({\mathfrak D}\nu) + \alpha_6({\mathfrak D}T) + T\beta_6}{T^2}\right]\\
&+\left(\kappa_9 - \frac{Q\mathfrak h_9}{E +P}\right) -\left( \frac{\mathfrak h  }{E + P}\right)\left[\frac{\alpha+\beta}{T} - \frac{\partial\beta}{\partial T} \right] + \left[\chi\left(\frac{\partial\nu}{\partial Q}\right) +\frac{\alpha}{T^2}\left(\frac{\partial T}{\partial Q}\right)\right]\left(\frac{\partial \kappa}{\partial T}\right)\\
s_7 =~& \left[\chi\left(\frac{\partial\nu}{\partial s}\right) +\frac{\alpha}{T^2}\left(\frac{\partial T}{\partial s}\right)\right]\frac{\partial}{\partial \nu}\left(\nu \mathfrak h-\kappa T\right)
-\left[\frac{T^2\chi_7({\mathfrak D}\nu) + \alpha_7({\mathfrak D}T) + T\beta_7}{T^2}\right]\\
&+\left(\kappa_8 - \frac{Q\mathfrak h_8}{E +P}\right) -\left( \frac{\mathfrak h  }{E + P}\right)\left[T\chi - \frac{\partial\beta}{\partial \nu} \right]+ \left[\chi\left(\frac{\partial\nu}{\partial Q}\right) +\frac{\alpha}{T^2}\left(\frac{\partial T}{\partial Q}\right)\right]\left(\frac{\partial \kappa}{\partial \nu}\right)\\
\end{split}
\end{equation*} 

\begin{equation}\label{app:eqt1}
\begin{split}
\mathfrak t_4=&-\frac{\tau_4}{T} +\left(\kappa_5 - \frac{Q\mathfrak h_5}{E +P}\right)+\left( \frac{\mathfrak h^2  }{E + P}\right)\\
\mathfrak t_5=&-\frac{\tau_5}{T} +\left(\kappa_6 - \frac{Q\mathfrak h_6}{E +P}\right)+\left( \frac{\mathfrak h  }{E + P}\right)\left(\frac{\partial\eta}{\partial T}\right)\\
\mathfrak t_6=&-\frac{\tau_6}{T} +\left(\kappa_7 - \frac{Q\mathfrak h_7}{E +P}\right)+\left( \frac{\mathfrak h  }{E + P}\right)\left(\frac{\partial\eta}{\partial \nu}\right)\\
\mathfrak t_7=&-\frac{\tau_7}{T},~~\mathfrak t_8=-\frac{\tau_8}{T},~~\mathfrak t_9=-\frac{\tau_9}{T}\\
\end{split}
\end{equation}

In deriving equations \eqref{app:eq1}, \eqref{app:eq2}, \eqref{app:eq3} and \eqref{app:eqt1} we have used the equation of motion expanded upto 2nd order in derivative expansion as described in equations \eqref{eomdetail} and \eqref{eomdetail2}. 
 \begin{equation}\label{eomdetail}
 \begin{split}
 (u.\partial s) + s\Theta
&=\left(\kappa - \frac{{Q\mathfrak h}}{E + P}\right) V^2 - \frac{\eta}{T}\sigma^2 + \left[\nu\left(\chi -{\mathfrak D}\chi\right) -\frac{(\alpha -{\mathfrak D}\alpha)}{T} - \frac{\beta}{T} \right]\Theta^2\\
 &+\left(\nu\chi - \frac{\alpha}{T}\right)(u.\partial\Theta) + \left(\nu\kappa - \frac{{\mathfrak h}}{T}\right)(D.V) + V^\mu D_\mu\left(\nu\kappa -\frac{{\mathfrak h} }{T}\right)\\
 &+\text{higher order terms}\\
 (u.\partial Q) + Q\Theta
&= -\left[\chi - {\mathfrak D}\chi\right]\Theta^2
 -\chi(u.\partial\Theta) -\kappa(D.V) - V^\mu D_\mu \kappa
 +\text{higher order terms}\\
 \end{split}
 \end{equation}
 
 \begin{equation}\label{eomdetail2}
 \begin{split}
&\left(E + P\right) h^\mu - TQ V^\mu \\
= ~&T\left[ \chi  - \frac{Q (\alpha + \beta)}{E + P} + \frac{3{\mathfrak D}{\mathfrak h}-4{\mathfrak h}}{3T} \right]\Theta V^\mu- {\mathfrak h}\sigma^{\mu\nu} V_\nu 
-P^{\mu\alpha}\left[ \beta D_\alpha\Theta + {\mathfrak h}(u.D)V_{\alpha}+\eta D_\nu\sigma^\nu_\alpha\right]\\
&+P^{\mu\alpha}\left[\left(T\chi -\frac{\partial\beta}{\partial\nu}\right)\Theta D_\alpha\nu + \left(\frac{\alpha +\beta}{T} - \frac{\partial\beta}{\partial T}\right)\Theta D_\alpha T  + V^\nu\left({\mathfrak h} \omega_{\alpha\nu} + \kappa F_{\alpha\nu}\right)\right]\\
&-\sigma^{\mu\nu}\left[\frac{\partial\eta}{\partial T} D_\nu T + \frac{\partial\eta}{\partial \nu} D_\nu \nu\right] + \text{higher order terms}
  \end{split}
 \end{equation}

\section{Notation}\label{app:notation}
In this section we shall summarize the notations we have used in various parts of this note. This might be useful since many similar looking symbols have been used to denote slightly different concepts or variables.
\begin{equation}\label{not:metric}
\begin{split}
\mathcal G_{\mu\nu} &= \text{Metric with arbitrary (but slow) space and time dependence}\\
G_{\mu\nu} &= \text{Most general static metric}\\
g_{ij} &= \text{Static metric on spatial slices}\\
D_\mu &= \text{Covariant derivative w.r.t. $\mathcal G_{\mu\nu}$}\\ 
\nabla_i &=\text{Covariant derivative w.r.t. $g_{ij}$}
\end{split}
\end{equation}

\begin{equation}\label{not:curve}
\begin{split}
&\tilde R = \text{Ricci scalar for $\mathcal G_{\mu\nu}$},~~\tilde R_{\mu\nu} = \text{Ricci tensor for $\mathcal G_{\mu\nu}$}\\
& R = \text{Ricci scalar for $g_{ij}$},~~R_{ij} = \text{Ricci tensor for $g_{ij}$}\\
\end{split}
\end{equation}

\begin{equation}\label{not:gauge}
\begin{split}
&T_0 = \text{Length of the time circle in static situation}\\
&\mathfrak A_\mu = \text{Gauge field with arbitrary space and time dependence}\\
&{\cal A}_i = \text{Space component of the static gauge field}\\
&A_0 = \text{Time component of the static gauge field}\\
&a_i =\left(\frac{G_{ti}}{G_{tt}}\right),~~\hat a_i = T_0 a_i,~~A_i = {\cal A}_i + A_0 a_i
\end{split}
\end{equation}

\begin{equation}\label{not:fluid}
\begin{split}
&T_0 = \text{Length of the time circle in static situation}\\
&\hat u^\mu =\left( \frac{\{1,0,0,0\}}{\sqrt{-G_{tt}}}\right),~~\hat T = \left( \frac{T_0}{\sqrt{-G_{tt}}}\right),~~\hat \nu = \frac{A_0}{T_0}\\
&u^\mu = \text{Fluid velocity},~~T = \text{ Temperature},~~\nu = \frac{\text{Chemical potential}}{\text{ Temperature}}\\
\end{split}
\end{equation}

\begin{equation}\label{not:fluidvari}
 \begin{split}
& P_{\mu\nu} = u_\mu u_\nu + {\mathcal G}_{\mu\nu}~ \text{where }~{\mathcal G}_{\mu\nu} = \text{metric}\\
&\Theta =D_\mu u^\mu,~~{\mathfrak a}_\mu = (u.D) u_\mu,~~h_\mu = \mathfrak a_\mu + P_\mu ^\alpha\left(\frac{ D_\alpha T}{T}\right)\\
&\omega_{\mu\nu} = P_\mu^\alpha P_\nu^\beta \left(\frac{D_\alpha u_\beta - D_\beta u_\alpha}{2}\right),~~\sigma_{\mu\nu} = P_\mu^\alpha P_\nu^\beta \left(\frac{D_\alpha u_\beta + D_\beta u_\alpha}{2}-\frac{\Theta}{3}{\mathcal G}_{\alpha\beta}\right)\\
 \end{split}
 \end{equation}
\begin{equation}\label{not:gauge2}
\begin{split}
&F_{\mu\nu} = \partial_\mu \mathfrak A_\nu - \partial_\nu \mathfrak A_\mu= \text{Field  strength for $\mathfrak A_\mu$}\\
&\hat F_{ij} = \partial _i A_j - \partial_j A_i= \text{Field  strength for $A_i$}\\
&\mathfrak f_{\mu\nu} = \partial_\mu u_\nu - \partial_\nu u_\mu\sim\text{Field  strength for $u_\mu$}\\
& f_{ij} = \partial _i  a_j - \partial_j  a_i= \text{Field  strength for $ a_i$}\\
&\hat f_{ij} = \partial _i \hat a_j - \partial_j \hat a_i= \text{Field  strength for $\hat a_i$}\\
&\bar F_{\mu\nu}=P_\mu^\alpha P_\nu^\beta F_{\alpha\beta},~\omega_{\mu\nu}=2P_\mu^\alpha P_\nu^\beta \mathfrak f_{\alpha\beta},~H_{\mu\nu} = \bar F_{\mu\nu} + 2 T \nu\omega_{\mu\nu}\\
&E_\mu = u^\nu F_{\mu\nu},~V_\mu  = \frac{E_\mu}{T} - P_\mu^\alpha D_\alpha \nu\\
\end{split}
\end{equation}
For any function $\mathfrak B$, dependent on  fluid variables and their derivatives, we have a corresponding symbol $\hat {\mathfrak B}$ which denotes the same quantity evaluated on $\{\hat u^\mu, \hat T,\hat\nu\}$ and in static background $G_{\mu\nu}$ and $\{A_0,\mathcal A_i\}$.
\newline
By $\mathfrak D M$ we denote the following.
\begin{equation}\label{notatiof1}
 \begin{split}
{\mathfrak D}M = s\left[\frac{\partial M}{\partial s}\right]_Q +  Q\left[\frac{\partial M}{\partial Q}\right]_s~\text{For any scalar $M(s,Q)$}
  \end{split}
 \end{equation}
 
\begin{equation}\label{not:current}
\begin{split}
&C^\mu = \text{Charge current},~~
T^{\mu\nu} = \text{Stress tensor},~~J^\mu = \text{Entropy current}\\
&j^\mu = \text{Derivative correction to the charge current}\\
&\pi^{\mu\nu} = \text{Derivative correction to the stress tensor}\\
\end{split}
\end{equation}


\providecommand{\href}[2]{#2}\begingroup\raggedright\endgroup
\end{document}